\documentclass[12pt,a4paper]{article}
\usepackage{amssymb,amsmath}
\newcommand{\ket}[1]{\left| #1 \right>} 
\newcommand{\bra}[1]{\left< #1 \right|} 
\newcommand{\braket}[2]{\left< #1 \vphantom{#2} \right|
 \left. #2 \vphantom{#1} \right>} 
\usepackage{color}
\usepackage{tikz}
\usepackage[T1]{fontenc}
\usepackage[utf8]{inputenc}
\usepackage{authblk}
\usepackage[title,titletoc,toc]{appendix}
\usepackage{adjustbox}
\usepackage{float}
\numberwithin{equation}{section}

\title{On relations between one-dimensional quantum and two-dimensional classical spin systems}
\author[1]{J. Hutchinson\footnote{j.hutchinson@bris.ac.uk}}
\author[1]{J.P. Keating\footnote{j.p.keating@bris.ac.uk}}
\author[1]{F. Mezzadri\footnote{f.mezzadri@bris.ac.uk}}
\affil[1]{School of Mathematics, University of Bristol, Bristol BS8 1TW, UK}

\begin{document}

\maketitle

\begin{abstract}
We exploit mappings between quantum and classical systems in order to obtain a class of two-dimensional classical systems with critical properties equivalent to those of the class of one-dimensional quantum systems discussed in a companion paper \cite{ours}. In particular, we use three approaches: the Trotter-Suzuki mapping; the method of coherent states; and a calculation based on commuting the quantum Hamiltonian with the transfer matrix of a classical system. This enables us to establish universality of certain critical phenomena by extension from the results in \cite{ours} for the classical systems identified.
\end{abstract}

\section{Introduction}
Mappings between statistical mechanical models have provided new pathways to compute thermodynamic properties of systems which were previously intractable~\cite{baxxyi,xyi,ts}. In particular, critical phenomena in $d$-dimensional quantum systems have been investigated by mapping them to $(d+1)$-dimensional classical systems for which there are better developed techniques, such as Monte Carlo simulations \cite{mc1,mc2}. For example, one well known connection is that between the one-dimensional XYZ model and the two-dimensional zero-field eight-vertex model, namely that the Hamiltonian of the quantum model and the transfer matrix of the classical model have the same eigenvectors. Baxter \cite{baxxyi} found the ground state energy for the XYZ model by first finding the partition function of the eight-vertex model and then showing that the quantum Hamiltonian is effectively the logarithmic derivative of the transfer matrix for the classical system.

In this paper we exploit these quantum to classical (QC) mappings for the opposite reason: to take advantage of known ground state critical behaviour in a general class of quantum spin chains to determine the finite-temperature critical properties of an equivalent class of classical spin systems.

In \cite{ours} we computed the critical exponents $s$, $\nu$ and $z$, corresponding to the energy gap, correlation length and dynamic exponent respectively, for a class of quantum spin chains, establishing universality for this class of systems. We also computed the ground state correlators $\left\langle  \sigma^{x}_{i} \sigma^{x}_{i+r} \right\rangle_{g}$,  $\left\langle \sigma^{y}_{i} \sigma^{y}_{i+r} \right\rangle_{g}$ and  $\left\langle \prod^{r}_{i=1} \sigma^{z}_{i} \right\rangle_{g}$ for this class of systems when translation invariance is imposed. These correlators were found to exhibit quasi-long range order behaviour when the systems are gapless, with a critical exponent dependent upon the system parameters. 

The class of quantum spin chains studied in \cite{ours} consists of $M$ spin-1/2 particles in an external field $h$, with a Hamiltonian quadratic in Fermi operators given by
\begin{equation}
\mathcal{H} = \sum^{M}_{j,k} \left( A_{jk} b^{\dagger}_{j} b_{k} + \frac{\gamma}{2} B_{jk} \left( b^{\dagger}_{j} b^{\dagger}_{k}- b_{j} b_{k} \right) \right) -   2 h \sum^{M}_{j=1} b^{\dagger}_{j} b_{j},
\label{generalh}
\end{equation}
where the $b_{j}$s are the Fermi operators satisfying the usual Fermi commutation relations 
\begin{equation}
\left\{b^{\dagger}_{j}, b_{k}\right\} = \delta_{j,k}, \quad \left\{b^{\dagger}_{j},b^{\dagger}_{k}\right\} = \left\{b_{j},b_{k}\right\} = \left(b^{\dagger}_{j} \right)^{2} = \left(b_{j}\right)^{2} = 0.
\label{fermicomm}
\end{equation}
The measure of anisotropy $\gamma$ is real, with $0 \leq \gamma \leq 1$; the matrix $A_{j,k}$ must be Hermitian and $B_{j,k}$ antisymmetric, both containing only real entries without loss of generality; and periodic boundary conditions $b_{M+j} = b_{j}$ are assumed.

This model can be diagonalised \cite{lieb} so that
\begin{equation}
\mathcal{H} =  \sum_{q} \left| \Lambda_{q} \right| \eta^{\dagger}_{q} \eta_{q} + C,
\label{hdiag}
\end{equation}
with the dispersion relation $\left| \Lambda_{q} \right|$ determined by matrices $A_{j,k}$ and $B_{j,k}$, the $\eta_{q}$s are Fermi operators, and $C$ is a constant.

In \cite{kmlong,kmshort} Keating and Mezzadri restricted the Hamiltonian \eqref{generalh} to possess symmetries corresponding to the Haar measure of each of the classical compact groups $U(N)$, $O^{+}(2N)$, $Sp(2N)$, $O^{+}(2N+1)$, $O^{-}(2N+1)$ and $O^{-}(2N+2)$, enabling the calculation of $\left| \Lambda_{q} \right|$ using techniques from random matrix theory. This corresponds to a symmetry classification of spin chains similar to that introduced for disordered systems by Atland and Zirnbauer \cite{altzin1,altzin2,zin3}. These symmetry properties were encoded into the structure of the matrices $A_{j,k}$ and $B_{j,k}$, as summarised in Table~\ref{tab:ccgg} in Appendix~\ref{sec:appsymcl}. For example, when restricted to $U(N)$ symmetry\footnote{For the other symmetry classes see \cite{kmlong}.} \cite{kmlong,kmshort}
\begin{equation}
\begin{aligned}
\Lambda_{q} & = 4  \left( \Gamma + \sum^{L}_{k=1} \left( a(k) \cos k q + i b(k) \sin k q \right) \right) \\
& = 4 \left( a_{q} + i b_{q} \right),
\label{lambda}
\end{aligned}
\end{equation}
with real and imaginary parts given by 
\begin{equation}
a_{q} = \Gamma + \sum^{L}_{k=1} a(k) \cos k q, \quad \mbox{and} \quad b_{q} = \sum^{L}_{k=1} b(k) \sin k q,
\label{aborig}
\end{equation}
and 
\begin{equation} 
\Gamma  = \frac{1}{2}
\begin{cases}
a(0), \quad L = \frac{M-1}{2}, & \mbox{if M is odd}, \\
a(0) + (-1)^{l} a(\frac{M}{2}) , \quad L = \frac{M}{2}-1, & \mbox{if M is even}, \\
\end{cases}
\end{equation}
where $q$ is the wave number
\begin{equation}
q = \frac{2 \pi l}{M} \quad \mbox{with} \quad l = 0, \ldots, M-1.
\end{equation}

In general, the symmetry constraints were achieved using real functions $a(j)$ and $b(j)$, even and odd functions of $\mathbb{Z}/M\mathbb{Z}$ respectively, to dictate the entries of matrices $A_{j,k}$ and $B_{j,k}$, as reported in Table~\ref{tab:ccgg} in Appendix~\ref{sec:appsymcl}.

Exploiting the formalism developed in \cite{kmlong,kmshort} enabled us to compute the critical properties of this class of spin chains \cite{ours}, demonstrating a dependence of the critical exponents on system symmetries and establishing universality for this class of quantum systems.

Having established universality for the above class of quantum spin chains in \cite{ours}, we now make use of QC mappings to obtain a class of classical systems with equivalent critical properties, establishing universality for this class of classical systems as well by extension. This is our main goal. 

A quantum and a classical system are equivalent if their partition functions are the same; such a correspondence, however, is not unique as different classical systems can be equivalent to the same quantum system. We shall here adopt the following different approaches to map the partition functions of the quantum spin chains~\eqref{generalh} onto those of a general class of two-dimensional classical systems:
\begin{itemize}
\item the Trotter-Suzuki formula (Section~\ref{sec:tsmap});
\item the method of coherent states (Section~\ref{sec:coh});
\item the simultaneous diagonalisation of the quantum Hamiltonian and the transfer matrix for the classical system (Section~\ref{sec:simdiag}).
\end{itemize}

\section{Trotter-Suzuki mapping}
\label{sec:tsmap}

This approach was developed by Suzuki \cite{ts}, who applied the Trotter product formula
\begin{equation}
e^{\hat{A}+ \hat{B}} = \left( e^{\frac{\hat{A}}{n}} e^{\frac{\hat{B}}{n}} \right)^{n}, \qquad \left[\hat{A}, \hat{B} \right] \neq 0,
\label{tpf}
\end{equation}
to map the partition function for a $d$-dimensional quantum system to that for a $(d+1)$-dimensional classical one. In particular he applied it to the partition function of a $d$-dimensional quantum Ising model in a transverse magnetic field, mapping it to that of a $(d+1)$-dimensional classical Ising model \cite{ts}. He then proved the equivalence of the critical properties of the ground state of the quantum system and the finite temperature properties of the classical system. 

Here we harness this technique to supply us with a class of two-dimensional classical systems with critical properties equivalent to those of the ground state of the quantum spin chains \eqref{generalh}. Like the original quantum system, the classical counterparts are also able to possess symmetries reflected by those of the Haar measure of each of the different classical compact groups\footnote{This is observed through the structure of matrices$A_{j,k}$ and $B_{j,k}$ summarised in Table~\ref{tab:ccgg} inherited by the classical systems.}, enabling the dependence of critical properties on system symmetries to be observed.

There are many ways to apply the Trotter-Suzuki mapping to the partition function for the class of quantum spin chains \eqref{generalh}, resulting in different classical partition functions.  Those that we obtain are of the form
\begin{subequations}
\label{zts1}
\begin{align}
\label{A1}
Z_A & = \sum_{\mbox{\tiny all states}} e^{-\beta_{\mbox{\tiny cl}} \mathcal{H}_{\mbox{\tiny cl}} \left( \left\{ s_{i,j} \right\} \right)} f\left( \left\{ s_{i,j} \right\} \right)  \\
\label{A2}
Z_B & = \sum_{\mbox{\tiny restricted states}} e^{ -\beta_{\mbox{\tiny cl}} \mathcal{H}_{\mbox{\tiny cl}}  \left( \left\{ s_{i,j} \right\} \right)} \\
\label{A3}
Z_C & = \sum_{\mbox{\tiny all configurations}} \prod_{i} \omega_{i} \\
\label{A4}
Z_D & = \sum_{\mbox{\tiny all states}} e^{-\beta_{\mbox{\tiny cl}} \mathcal{H}_{\mbox{\tiny cl}}  \left( \left\{ \sigma_{i,j} \right\}, \left\{ \tau_{i,j} \right\}, \left\{ s_{i,j} \right\} \right)}
\end{align}
\end{subequations}
where $\mathcal{H}_{\mbox{\tiny cl}}$ is the effective classical Hamiltonian. In \eqref{A1} and~\eqref{A2}
$\mathcal{H}_{\mbox{\tiny cl}}$ is a real function of the classical spin variables $s_{i,j}=\pm 1$ and in \eqref{A4} it is a complex function of the classical spin variables $\sigma_{i,j},\tau_{i,j},s_{i,j} = \pm 1$, which represent the eigenvalues of the Pauli matrices $\sigma^{x}_{i}$, $\sigma^{y}_{i}$ and $\sigma^{z}_{i}$ respectively. The function $f\left( \left\{s_{i,j} \right\} \right)$ is also a real function of the classical spin variables $s_{i,j} = \pm 1$, and we find that if $f\left( \left\{ s_{i,j} \right\} \right) = 1$, then \eqref{A1} has the familiar form of a classical partition function, with $\mathcal{H}_{\mbox{\tiny cl}}$ representing the Hamiltonian describing the effective classical system. The same is true for \eqref{A2} and \eqref{A4}, but \eqref{A2} has additional constraints on the spin states and \eqref{A4} involves imaginary interaction coefficients. The form in \eqref{A3} is that of a vertex model with vertex weights given by $\omega_{i}$.


We begin to present our results by first restricting to quantum systems with  nearest neighbour interactions only. The extensions to longer-range interactions are detailed in Appendix~\ref{sec:lri}. 

\subsection{Nearest neighbour interactions }
\label{sec:nn}

Restricting \eqref{generalh} to nearest-neighbour interactions gives the well known one-dimensional quantum XY model\footnote{We can ignore boundary term effects since we are interested in the thermodynamic limit only.}  
\begin{equation}
\mathcal{H}^{XY} = - \frac{1}{2} \sum^{M}_{j=1}   \left( J^{x}_{j} \sigma^{x}_{j}  \sigma^{x}_{j+1} + J^{y}_{j} \sigma^{y}_{j}  \sigma^{y}_{j+1} + h \sigma^{z}_{j} \right),
\label{xymodel}
\end{equation}
where $J^{x}_{j}= - \left(A_{j,j+1} +\gamma B_{j,j+1} \right)$, $J^{y}= -\left(A_{j,j+1} -\gamma B_{j,j+1} \right)$. This mapping is achieved by using Jordan-Wigner transformations:
\begin{equation}
\begin{aligned}
b^{\dagger}_{j} & = \frac{1}{2} \left( m_{2j+1} + i m_{2j} \right) \\
& = \frac{1}{2} \left(\sigma^{x}_{j} + i \sigma^{y}_{j} \right) \prod^{j-1}_{l=1} \left(- \sigma^{z}\right),
\end{aligned}
\quad
\begin{aligned}
b_{j} & = \frac{1}{2} \left( m_{2j+1} - i m_{2j} \right) \\
& = \frac{1}{2} \left(\sigma^{x}_{j} - i \sigma^{y}_{j} \right) \prod^{j-1}_{l=1} \left(- \sigma^{z}_{l}\right),
\label{mbtrans}
\end{aligned}
\end{equation}
where
\begin{equation}
m_{2j+1} = \sigma^{x}_{j} \prod^{j-1}_{l=0} \left(- \sigma^{z}_{j}\right), \quad m_{2j} = \sigma^{y}_{j} \prod^{j-1}_{l=0} \left(-\sigma^{z}_{j}\right),
\label{minustrans}
\end{equation}
or inversely as
\begin{equation}
\begin{aligned}
& \sigma^{z}_{j} = i m_{2j} m_{2j+1}, \quad \sigma^{x}_{j} = m_{j2+1} \prod^{j-1}_{l=0} \left( - i m_{2l} m_{2l+1} \right) , \\
& \mbox{and} \quad \sigma^{y}_{j} = m_{2j} \prod^{j-1}_{l=0} \left(- i m_{2l} m_{2l+1} \right).
\end{aligned}
\end{equation}
The $m_{j}$s are thus Hermitian and obey the anti-commutation relations $\left\{m_{j}, m_{k}\right\} = 2 \delta_{jk}$.

\subsubsection{A class of classical Ising type models \eqref{A1}}

When we restrict to $\gamma = 1$ and $B_{j,j+1} = A_{j,j+1}$, \eqref{xymodel} becomes a class of quantum Ising type models in a transverse magnetic field with site-dependent coupling parameters. Suzuki showed~\cite{ts} that the partition function for such a system can be mapped\footnote{Upto an overall constant.} to that for a class of two-dimensional classical Ising models with Hamiltonian $\mathcal{H}_{\mbox{\tiny cl}}$ given by
\begin{equation}
\mathcal{H}_{\mbox{\tiny cl}} = -\sum^{n}_{p=1} \sum^{M}_{j=1} \left( J^{h}_{j} s_{j,p} s_{j+1,p} + J^{v} s_{j,p} s_{j,p+1}  \right),
\label{clis}
\end{equation}
with parameter relations
\begin{equation}
\beta_{\mbox{\tiny cl}}J^{v} = \frac{1}{2} \log{\coth{\frac{\beta_{\mbox{\tiny qu}} h}{n}}}, \quad \beta_{\mbox{\tiny cl}} J^{h}_{j} = \frac{\beta_{\mbox{\tiny qu}}}{n} J_{j}, 
\label{paramrelis}
\end{equation} 
where $\beta_{\mbox{\tiny qu(cl)}}$ is the inverse temperature of the quantum (classical) system.

Thus we have an equivalence between our class of quantum spin chains under these restrictions and a class of two-dimensional classical Ising models also with site-dependent coupling parameters in one direction and a constant coupling parameter in the other. From \eqref{paramrelis} we see that the magnetic field $h$ driving the phase transition in the ground state of the quantum system plays the role of temperature $\beta_{\mbox{\tiny cl}}$ driving the finite temperature phase transition of the classical system.

This mapping holds in the limit $n \rightarrow \infty$, which would result in anisotropic couplings for the class of classical Ising models, unless we also take $\beta_{\mbox{\tiny qu}} \rightarrow \infty$. This therefore provides us with a connection between the ground state properties of the class of quantum systems and the finite temperature properties of the classical. 

In this case we can also use this mapping to write the expectation value of any function $f\left(\left\{\sigma^{z}\right\}\right)$ with respect to the ground state of the class of quantum systems as
\begin{equation}
\left\langle f\left(\left\{\sigma^{z}\right\}\right) \right\rangle_{\mbox{\tiny qu}} =  \left\langle f\left(\left\{ s \right\}\right) \right\rangle_{\mbox{\tiny cl}},
\label{expec}
\end{equation}
where $\left\langle f\left(\left\{ s \right\}\right) \right\rangle_{\mbox{\tiny cl}}$ is the finite temperature expectation of the corresponding function of classical spin variables with respect to the class of classical systems \eqref{clis}. 

Some examples of this are the spin correlation functions between two or more spins in the ground state of the class of quantum systems in the $z$ direction, which can be interpreted as the equivalent correlator between classical spins in the same row of the corresponding class of classical systems \eqref{clis};
\begin{equation}
\left\langle \sigma^{z}_{j} \sigma^{z}_{j+r} \right\rangle_{\mbox{\tiny qu}}  = \left\langle s_{j,p} s_{j+r,p} \right\rangle_{\mbox{\tiny cl}}, \qquad \left\langle \prod^{r}_{j=1} \sigma^{z}_{j} \right\rangle_{\mbox{\tiny qu}}  = \left\langle \prod^{r}_{j=1} s_{j,p} \right\rangle_{\mbox{\tiny cl}}.
\label{equivcorr}
\end{equation}

\subsubsection{A class of classical Ising type models with additional constraints on the spin states \eqref{A2}}
\label{sec:conds}

Similarly, the Trotter Suzuki mapping can be applied to the partition function for the XY model \eqref{xymodel} in full generality. In this case we first order the terms in the partition function in the following way 
\begin{equation}
\begin{aligned}
Z & = \lim_{n \rightarrow \infty} \mbox{Tr } \left[ \hat{\mathcal{V}}_{a} \hat{\mathcal{V}}_{b} \right]^{n}, \quad  \hat{\mathcal{V}}_{\alpha} = \sum_{j \in \alpha} e^{\frac{\beta_{\mbox{\tiny qu}}}{n}\hat{\mathcal{H}}^{z}_{j}} e^{\frac{\beta_{\mbox{\tiny qu}}}{n}\hat{\mathcal{H}}^{x}_{j}} e^{\frac{\beta_{\mbox{\tiny qu}}}{n} \hat{\mathcal{H}}^{y}_{j}} e^{\frac{\beta_{\mbox{\tiny qu}}}{n} \hat{\mathcal{H}}^{z}_{j}},
\label{vmxyid}
\end{aligned}
\end{equation}
where $\hat{\mathcal{H}}^{\mu}_{j} = J^{\mu}_{j} \sigma^{\mu}_{j} \sigma^{\mu}_{j+1}$ for $\mu \in x,y$, $\hat{\mathcal{H}}^{z}_{j} = \frac{h}{4} \left( \sigma^{z}_{j} + \sigma^{z}_{j+1} \right)$, and $\alpha$ denotes either $a$ or $b$, which are the sets of odd and even integers respectively. 

We then insert $2 n$ copies of the identity operator in the $\sigma^{z}$ basis; $\mathbb{I}_{s_{p}} = \sum_{s} \ket{\vec{s}_{p}} \bra{\vec{s}_{p}}$ where $\ket{\vec{s}_{p}}= \ket{s_{1p},s_{2p},\ldots,s_{Mp}}$ between each of the $2n$ terms in \eqref{vmxyid};
\begin{equation}
\begin{aligned}
Z & = \lim_{n \rightarrow \infty} \mbox{Tr } \mathbb{I}_{s_{1}} \hat{\mathcal{V}}_{a} \mathbb{I}_{s_{2}} \hat{\mathcal{V}}_{b} \ldots \mathbb{I}_{s_{2n-1}} \hat{\mathcal{V}}_{a} \mathbb{I}_{s_{2n}} \hat{\mathcal{V}_{b}}\\
& = \lim_{n \rightarrow \infty} \sum_{s_{j,p}} \prod^{2n}_{p \in a} \bra{\vec{s}_{j,p}} \hat{\mathcal{V}}_{a} \ket{\vec{s}_{j,p+1}} \bra{\vec{s}_{j,p+1}} \hat{\mathcal{V}}_{b} \ket{\vec{s}_{j,p+2}}.
\label{zvv}
\end{aligned}
\end{equation}

The remaining matrix elements in \eqref{zvv} are given by
\begin{equation}
\begin{aligned}
\bra{\vec{s}_{j,p}} \hat{\mathcal{V}}_{\alpha} \ket{\vec{s}_{j,p+1}} & = \prod^{M}_{j \in \alpha} \bra{s_{j,p}, s_{j+1,p}} \mathbf{M} \ket{s_{j,p+1},s_{j+1,p+1}},
\label{me}
\end{aligned}
\end{equation}
where
\begin{equation}
\begin{aligned}
\mathbf{M} & = 
\begin{pmatrix}
\overset{\bra{\uparrow \uparrow} \mathbf{M} \ket{\uparrow \uparrow}}{e^{\frac{\beta_{\mbox{\tiny qu}} h}{n}} \cosh{\left( \frac{2 \beta_{\mbox{\tiny qu}} \gamma}{n} B_{j}\right)}} & 0 & 0 & \overset{\bra{\uparrow \uparrow} \mathbf{M} \ket{\downarrow \downarrow}}{\sinh{\left(\frac{ 2 \beta_{\mbox{\tiny qu}} \gamma}{n} B_{j}\right)}}  \\
0 & \overset{\bra{\uparrow \downarrow} \mathbf{M} \ket{\uparrow \downarrow}}{\cosh{\left(\frac{2 \beta_{\mbox{\tiny qu}}}{n} A_{j}\right)}} & \overset{\bra{\uparrow \downarrow} \mathbf{M} \ket{\downarrow \uparrow}}{\sinh{\left( \frac{2 \beta_{\mbox{\tiny qu}}}{n} A_{j} \right)}} & 0 \\
0 & \overset{\bra{\downarrow \uparrow} \mathbf{M} \ket{\uparrow \downarrow}}{\sinh{\left( \frac{2 \beta_{\mbox{\tiny qu}}}{n} A_{j} \right)}} & \overset{\bra{\downarrow \uparrow} \mathbf{M} \ket{\downarrow \uparrow}}{\cosh{\left( \frac{2 \beta_{\mbox{\tiny qu}}}{n} A_{j} \right)}} & 0  \\
\overset{\bra{\downarrow \downarrow} \mathbf{M} \ket{\uparrow \uparrow}}{\sinh{\left( \frac{ 2 \beta_{\mbox{\tiny qu}} \gamma}{n} B_{j}\right)}} & 0 & 0 & \overset{\bra{\downarrow \downarrow} \mathbf{M} \ket{\downarrow \downarrow}}{e^{-\frac{h}{n}} \cosh{\left(\frac{2 \beta_{\mbox{\tiny qu}} \gamma}{n} B_{j}\right)}}
\end{pmatrix}.
\label{xymatrix}
\end{aligned}
\end{equation}

\begin{figure}[H]
\begin{center}
\includegraphics[width=0.8\textwidth,height=0.45\textheight]{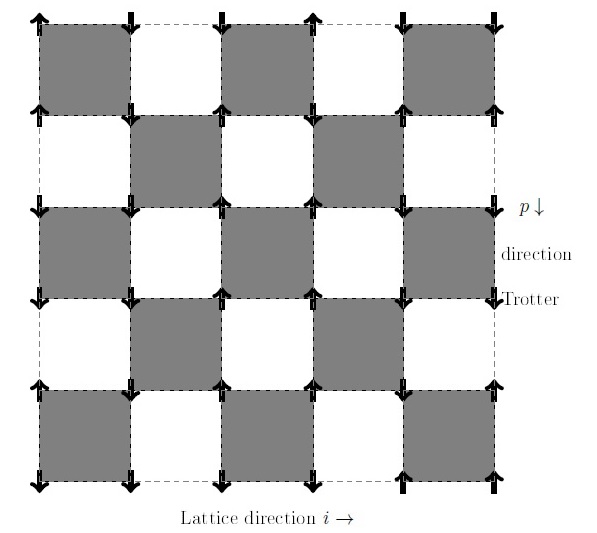}
\end{center}
\caption{Lattice representation of a class of classical systems equivalent to the general class of quantum systems \eqref{xymodel}. Spins only interact with other spins bordering the same shaded square.}
  \label{figure:shadedsqs}
\end{figure}

It is then possible to write the terms \eqref{me} in exponential form as
\begin{equation}
\begin{aligned}
\bra{\vec{s}_{j,p}} \hat{\mathcal{V}}_{\alpha} \ket{\vec{s}_{j,p+1}} & = \prod^{M}_{j \in \alpha} e^{- \beta_{\mbox{\tiny cl}} \mathcal{H}_{j,p}}, 
\label{expform}
\end{aligned}
\end{equation}
where $\mathcal{H}_{j,p}$ can be written as
\begin{equation}
\begin{aligned}
\mathcal{H}_{j,p} =  -\frac{1}{4} ( &J^{v}_{j}  s_{j,p} s_{j,p+1}  + J^{h}_{j} s_{j,p} s_{j+1,p} + J^{d}_{j} s_{j+1,p} s_{j,p+1} \\
& + H\left( s_{j,p} + s_{j+1,p} \right) + C_{j} ),
\label{hsymmless}
\end{aligned}
\end{equation}
or more symmetrically as
\begin{equation}
\begin{aligned}
\mathcal{H}_{j,p} =- \frac{1}{4} ( & J^{h}_{j} \left(s_{j,p}s_{j+1,p} + s_{j,p+1} s_{j+1,p+1} \right)  + J^{v}_{j} \left(s_{j,p} s_{j,p+1} + s_{j+1,p} s_{j+1,p+1} \right) \\
& + J^{d}_{j} \left( s_{j,p} s_{j+1,p+1} + s_{j,p+1} s_{j+1,p} \right)  \\
& + H \left( s_{j,p} + s_{j+1,p} + s_{j,p+1} + s_{j+1,p+1}\right) + C_{j}) ,
\label{hsymmmore}
\end{aligned}
\end{equation}
where 
\begin{equation}
\begin{aligned}
& \beta_{\mbox{\tiny cl}} J^{h}_{j} =  \log \frac{\sinh \frac{4 \beta_{\mbox{\tiny qu}}}{n} \gamma B_{j}}{\sinh \frac{ 4 \beta_{\mbox{\tiny qu}}}{n} A_{j}}, \quad \beta_{\mbox{\tiny cl}} J^{d}_{j} = \log \frac{\tanh \frac{2\beta_{\mbox{\tiny qu}}}{n} A_{j}}{\tanh \frac{2 \gamma \beta_{\mbox{\tiny qu}}}{n} B_{j}}, \\
&  \beta_{\mbox{\tiny cl}} J^{v}_{j} =   \log \coth \frac{2 \gamma \beta_{\mbox{\tiny qu}}}{n} B_{j} \coth \frac{2 \beta_{\mbox{\tiny qu}}}{n} A_{j},  \\
& \beta_{\mbox{\tiny cl}} H = \frac{\beta_{\mbox{\tiny qu}}h}{n}, \quad  \beta_{\mbox{\tiny cl}}  C_{j} = \log \sinh \frac{2 \beta_{\mbox{\tiny qu}}}{n} A_{j} \sinh \frac{2 \gamma \beta_{\mbox{\tiny qu}}}{n} B_{j},
\end{aligned}
\end{equation}
as long as we have the additional restriction that the four spins bordering each shaded square in Figure~\ref{figure:shadedsqs} obey
\begin{equation}
s_{j,p} s_{j+1,p} s_{j,p+1} s_{j+1,p+1} = 1.
\label{cond}
\end{equation}
This guarantees that each factor in the partition function is different from zero.

Thus we obtain a partition function equivalent to that for a class of two-dimensional classical Ising type models on a $M \times 2n$ lattice with classical Hamiltonian $\mathcal{H}_{\mbox{\tiny cl}}$ given by
\begin{equation}
\mathcal{H}_{\mbox{\tiny cl}} = \left(\sum^{2n}_{p \in a} \sum^{M}_{j \in a} + \sum^{2n}_{p \in b} \sum^{M}_{j \in b} \right) \mathcal{H}_{j,p},
\label{clnocond}
\end{equation} 
where $\mathcal{H}_{j,p}$ can have the form \eqref{hsymmless} or \eqref{hsymmmore}, with the additional constraint \eqref{cond}.

In this case we see that the classical spin variables at each site of the two-dimensional lattice only interact with other spins bordering the same shaded square, represented schematically in Figure~\ref{figure:shadedsqs}, with an even number of these four interacting spins being spin up and down (from condition \eqref{cond}). 

This mapping holds in the limit $n \rightarrow \infty$, which would result in coupling parameters $J^{h}_{j}, J^{d}_{i}, H \rightarrow 0$ and $J^{v}_{j} \rightarrow \infty$ unless we also take $\beta_{\mbox{\tiny qu}} \rightarrow \infty$. Therefore this again gives us a connection between the ground state properties of this class of quantum systems and the finite temperature properties of the classical systems. 

Again we have the same relationship between expectation values \eqref{expec} and \eqref{equivcorr}.

\subsection{A class of classical Ising type models with imaginary interaction coefficients \eqref{A4}}

Alternatively, lifting the restriction \eqref{cond} we instead can obtain a class of classical systems described by a Hamiltonian containing imaginary interaction coefficients
\begin{equation}
\mathcal{H}_{\mbox{\tiny cl}} = -\sum^{n}_{p=1} \sum^{M}_{j=1} \left(J^{\sigma}_{j} \sigma_{j,p} \sigma_{j+1,p} + J^{\tau}_{j} \tau_{j,p} \tau_{j+1,p} + i J \tau_{j,p} \left(  \sigma_{j,p} - \sigma_{j,p+1} \right) \right),
\label{clnnn}
\end{equation}
with parameter relations given by
\begin{equation}
\beta_{\mbox{\tiny cl}} J^{\sigma}_{j} = \frac{\beta_{\mbox{\tiny qu}}}{n} J^{x}_{j},  \quad \beta_{\mbox{\tiny cl}} J^{\tau}_{j} = \frac{\beta_{\mbox{\tiny qu}}}{n} J^{y}_{j}, \quad \beta_{\mbox{\tiny cl}} J = \frac{1}{2}\arctan{\frac{1}{\sinh{ \frac{ \beta_{\mbox{\tiny qu}} }{n}h }}}.
\end{equation}

To achieve this, we first apply the Trotter-Suzuki mapping to the quantum partition function divided in the following way
\begin{equation}
\begin{aligned}
Z = & \lim_{n \rightarrow \infty} \mbox{Tr } \left[ \hat{\mathcal{U}}_{1} \hat{\mathcal{U}}_{2} \right]^{n}, \\
\hat{\mathcal{U}}_{1} = &e^{\frac{\beta_{\mbox{\tiny qu}}}{2n} \hat{\mathcal{H}}_{x}} e^{\frac{\beta_{\mbox{\tiny qu}}}{2n} \hat{\mathcal{H}}_{z}} e^{\frac{\beta_{\mbox{\tiny qu}}}{2n} \hat{\mathcal{H}}_{y}} , \quad \hat{\mathcal{U}}_{2} = e^{\frac{\beta_{\mbox{\tiny qu}}}{2n} \hat{\mathcal{H}}_{y}} e^{\frac{\beta_{\mbox{\tiny qu}}}{2n} \hat{\mathcal{H}}_{z}} e^{\frac{\beta_{\mbox{\tiny qu}}}{2n} \hat{\mathcal{H}}_{x}}
\label{z1xy}
\end{aligned}
\end{equation}
where this time $\hat{\mathcal{H}}^{\mu} = \sum^{M}_{j=1} J^{\mu}_{j} \sigma^{\mu}_{j} \sigma^{\mu}_{j+1} $ for $\mu \in x,y$ and $\hat{\mathcal{H}}^{z} = \sum^{M}_{j=1}  \sigma^{z}_{j}$.

Next insert $n$ of each of the identity operators $\mathbb{I}_{\sigma_{p}}= \sum_{\sigma_{p}} \ket{\vec{\sigma}_{p}} \bra{\vec{\sigma}_{p}}$ and $\mathbb{I}_{\tau_{p}} = \sum_{\tau_{p}} \ket{\vec{\tau}_{p}} \bra{\vec{\tau}_{p}}$, which are in the $\sigma^{x}$ and $\sigma^{y}$ basis respectively, into~\eqref{z1xy} obtaining 
\begin{equation}
\begin{aligned}
Z  = \lim_{n \rightarrow \infty} & \mbox{Tr } \mathbb{I}_{\sigma_{1}} \hat{\mathcal{U}}_{1} \mathbb{I}_{\tau_{1}} \hat{\mathcal{U}}_{2} \mathbb{I}_{\sigma_{2}} \hat{\mathcal{U}}_{1} \mathbb{I}_{\tau_{2}} \ldots  \mathbb{I}_{\tau_{2n}} \hat{\mathcal{U}}_{2} \\
=  \lim_{n \rightarrow \infty} & \sum_{\sigma_{j,p}, \tau_{j,p}} \prod^{n}_{p=1} \bra{\vec{\sigma}_{p}} \hat{\mathcal{U}}_{1} \ket{\vec{\tau}_{p}} \bra{\vec{\tau}_{p}} \hat{\mathcal{U}}_{2} \ket{\vec{\sigma}_{p+1}}. 
\label{z2xy}
\end{aligned}
\end{equation}

It is then possible to rewrite the remaining matrix elements in \eqref{z2xy} as complex exponentials,
\begin{equation}
\begin{aligned}
\bra{\vec{\sigma}_{p}} \hat{\mathcal{U}}_{1}  \ket{\vec{\tau}_{p}}  & \bra{\vec{\tau}_{p}} \hat{\mathcal{U}}_{2} \ket{\vec{\sigma}_{p+1}} = e^{\frac{\beta_{\mbox{\tiny qu}}}{n} \left( \frac{1}{2} \left(\mathcal{H}^{x}_{p} + \mathcal{H}^{x}_{p+1} \right) + \mathcal{H}^{y}_{p} \right)} \\
& \qquad \times \bra{\vec{\sigma}_{p}} e^{\frac{\beta_{\mbox{\tiny qu}}}{2n}\hat{\mathcal{H}}^{z}} \ket{\vec{\tau}_{p}} \bra{\vec{\tau}_{p}} e^{\frac{\beta_{\mbox{\tiny qu}}}{2n} \hat{\mathcal{H}}^{z}} \ket{\vec{\sigma}_{p+1}}\\
& = C^{2M} e^{\frac{\beta_{\mbox{\tiny qu}}}{n} \left( \frac{1}{2} \left(\mathcal{H}^{x}_{p} + \mathcal{H}^{x}_{p+1} \right) + \mathcal{H}^{y}_{p} \right) +\frac{i}{2} D \sum^{M}_{j=1} \tau_{j,p} \left(  \sigma_{j,p} - \sigma_{j,p+1} \right)},
\end{aligned}
\end{equation}
where $\mathcal{H}^{x}_{p} = \sum^{M}_{j=1} J^{x} \sigma_{j,p} \sigma_{j+1,p}$, $\mathcal{H}^{y}_{p} = \sum^{M}_{j=1} J^{y} \tau_{j,p} \tau_{j+1,p}$, $D = \frac{1}{2}\arctan{\frac{1}{\sinh{ \frac{ \beta_{\mbox{\tiny qu}} }{n}h }}}$, \\
$C = \frac{1}{2} \cosh \left(\frac{\beta_{\mbox{\tiny qu}} }{n}h \right)$, and we have used 
\begin{equation}
\bra{\sigma_{j,p}} e^{ a\sigma^{z}_{j}} \ket{\tau_{j,p}} = \frac{1}{2} \cosh{\left(2a \right)} e^{i \frac{1}{2}\arctan{ \left(\frac{1}{\sinh{ \left(2a \right)}} \right)} \sigma_{j,p}\tau_{j,p} }.
\end{equation}

\begin{figure}[H]
\begin{center}
\includegraphics[width=1\textwidth,height=0.3\textheight]{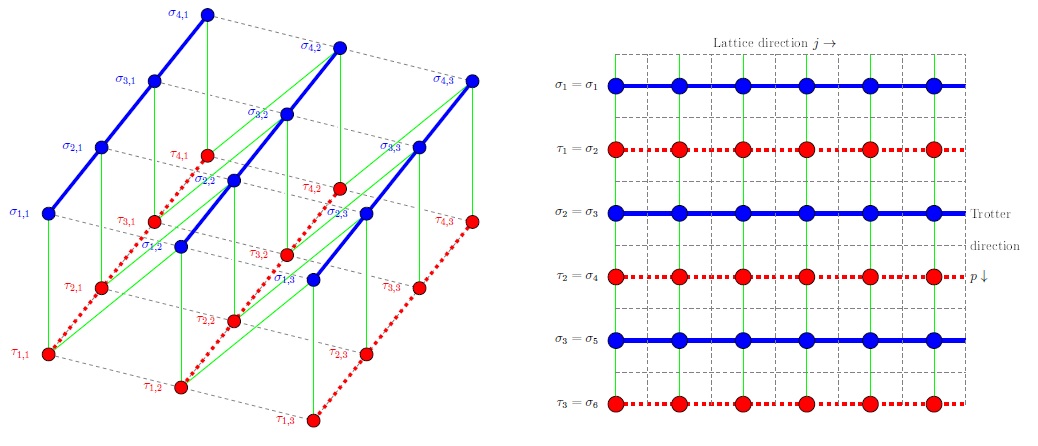}
\end{center}
\caption{Lattice representation of a class of classical systems equivalent to the class of quantum systems \eqref{xymodel}. The picture on the left can be ``unfolded'' into that on the right.}
\label{figure:xyim}
\end{figure}

The classical system with Hamiltonain given by \eqref{clnnn} can be depicted as in Figure~\ref{figure:xyim}, where the two types of classical spin variables $\sigma_{j,p}$ and $\tau_{j,p}$ can be visualised as each representing two-dimensional lattices on two separate planes, as shown on the left in Figure~\ref{figure:xyim}. The blue (thick solid) lines represent interactions with coefficients dictated by $J^{\sigma}_{j}$, the red (thick dashed) lines by $J^{\tau}_{j}$, and the $J_{i}$ coupling constants correspond to the green (thin solid) lines which connect these two lattice interaction planes. One can imagine ``unfolding'' the three-dimensional interaction surface shown on the left in Figure~\ref{figure:xyim} into the two-dimensional plane shown on the right.

As in previous cases, this mapping holds in the limit $n \rightarrow \infty$, which would result in coupling parameters $J^{\sigma}_{j}, J^{\tau}_{j} \rightarrow \infty$ and $J \rightarrow \frac{\pi}{4 \beta_{\mbox{\tiny cl}}}$ unless we also take $\beta_{\mbox{\tiny qu}} \rightarrow 0$. Therefore, it gives us a connection between the ground state properties of the class of quantum systems and the finite temperature properties of the classical ones. 

We can use this mapping to write the expectation value of any function $f\left(\left\{\sigma^{x}\right\}\right)$ or $f\left(\left\{\sigma^{y}\right\}\right)$ with respect to the ground state of the class of quantum systems \eqref{xymodel} as
\begin{equation}
\begin{aligned}
\left\langle f\left(\left\{\sigma^{x}\right\}\right) \right\rangle_{\mbox{\tiny qu}} = \left\langle f\left(\left\{\sigma\right\}\right) \right\rangle_{\mbox{\tiny cl}}, \qquad  \left\langle f\left(\left\{\sigma^{y}\right\}\right) \right\rangle_{\mbox{\tiny qu}} = \left\langle f\left(\left\{\tau\right\}\right) \right\rangle_{\mbox{\tiny cl}}, 
\label{fcxy}
\end{aligned}
\end{equation}
where $\left\langle f\left(\left\{\sigma\right\}\right) \right\rangle_{\mbox{\tiny cl}}$ and $ \left\langle f\left(\left\{\tau\right\}\right) \right\rangle_{\mbox{\tiny cl}}$ are the finite temperature expectation values of the equivalent function of classical spin variables with respect to the class of classical systems \eqref{clnnn}.\footnote{Recall from the picture on the right in Figure~\ref{figure:xyim}, that the $\sigma$ and $\tau$ represent alternate rows of the lattice.}

An example of this is the two-spin correlation function between spins in the ground state of the class of quantum systems \eqref{xymodel} in the $x$ and $y$ direction which can be interpreted as the two-spin correlation function between spins in the same odd and even rows of the corresponding class of classical systems \eqref{clis} respectively;
\begin{equation}
\left\langle \sigma^{x}_{j} \sigma^{x}_{j+r} \right\rangle_{\mbox{\tiny qu}}  =  \left\langle \sigma_{j,p} \sigma_{j+r,p} \right\rangle_{\mbox{\tiny cl}}, \qquad
\left\langle \sigma^{y}_{j} \sigma^{y}_{j+r} \right\rangle_{\mbox{\tiny qu}} = \left\langle \tau_{j,p} \tau_{j+r,p} \right\rangle_{\mbox{\tiny cl}}.
\label{qcy}
\end{equation}

\subsection{A class of classical vertex models \eqref{A3}}

Another interpretation of the partition function obtained using the Trotter Suzuki mapping, following a similar method to that of~\cite{barmash}, is that corresponding to a vertex model.

This can be seen by applying the Trotter Suzuki mapping to the quantum partition function ordered as in \eqref{vmxyid} and inserting $2n$ identity operators as in \eqref{zvv}, with remaining matrix elements given once more by \eqref{me}. This time, instead of writing them in exponential form as in \eqref{expform}, we interpret each matrix element as a weight corresponding to a different vertex configuration at every point $(j,p)$ of the lattice;
\begin{equation}
\begin{aligned}
\bra{\vec{s}_{j,p}} e^{\frac{\beta_{\mbox{\tiny qu}}}{n} \mathcal{V}_{\alpha}} \ket{\vec{s}_{j,p+1}} & = \prod^{M}_{j \in \alpha} \omega^{j} \left(s_{j,p},s_{j+1,p}, s_{j,p+1},s_{j+1,p+1} \right).
\end{aligned}
\end{equation}

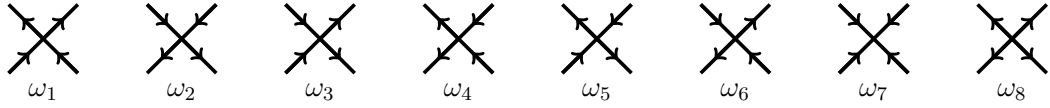
\begin{figure}[H]
  \centering
	\begin{adjustbox}{max size={1\textwidth}{1\textheight}}
  \begin{tikzpicture}
		
		\node [below] at (0.5,0) {$\omega_{1}$};
		\node [below] at (2.5,0) {$\omega_{2}$};
		\node [below] at (4.5,0) {$\omega_{3}$};
		\node [below] at (6.5,0) {$\omega_{4}$};
		\node [below] at (8.5,0) {$\omega_{5}$};
		\node [below] at (10.5,0) {$\omega_{6}$};
		\node [below] at (12.5,0) {$\omega_{7}$};
		\node [below] at (14.5,0) {$\omega_{8}$};

		
		\draw[->,line width=1.5] (0,0) to (0.3,0.3);
		\draw[->,line width=1.5] (0.5,0.5) to (0.2,0.8);
		\draw[->,line width=1.5] (0.5,0.5) to (0.8,0.8);
		\draw[->,line width=1.5] (1,0) to (0.7,0.3);
		
		\draw[->,line width=1.5] (2.5,0.5) to (2.2,0.2);
		\draw[->,line width=1.5] (2.5,0.5) to (2.8,0.2);
		\draw[->,line width=1.5] (2,1) to (2.3,0.7);
		\draw[->,line width=1.5] (3,1) to (2.7,0.7);
		
		\draw[->,line width=1.5] (4,1) to (4.3,0.7);
		\draw[->,line width=1.5] (4.5,0.5) to (4.8,0.8);
		\draw[->,line width=1.5] (4,0) to (4.3,0.3);
		\draw[->,line width=1.5] (4.5,0.5) to (4.8,0.2);
		
		\draw[->,line width=1.5] (6.5,0.5) to (6.2,0.8);
		\draw[->,line width=1.5] (7,1) to (6.7,0.7);
		\draw[->,line width=1.5] (6.5,0.5) to (6.2,0.2);
		\draw[->,line width=1.5] (7,0) to (6.7,0.3);
		
		\draw[->,line width=1.5] (8.5,0.5) to (8.2,0.8);
		\draw[->,line width=1.5] (9,1) to (8.7,0.7);
		\draw[->,line width=1.5] (8,0) to (8.3,0.3);
		\draw[->,line width=1.5] (8.5,0.5) to (8.8,0.2);
		
		\draw[->,line width=1.5] (10,1) to (10.3,0.7);
		\draw[->,line width=1.5] (10.5,0.5) to (10.8,0.8);
		\draw[->,line width=1.5] (10.5,0.5) to (10.2,0.2);
		\draw[->,line width=1.5] (11,0) to (10.7,0.3);
		
		\draw[->,line width=1.5] (12,1) to (12.3,0.7);
		\draw[->,line width=1.5] (13,1) to (12.7,0.7);
		\draw[->,line width=1.5] (12,0) to (12.3,0.3);
		\draw[->,line width=1.5] (13,0) to (12.7,0.3);

		\draw[->,line width=1.5] (14.5,0.5) to (14.2,0.2);
		\draw[->,line width=1.5] (14.5,0.5) to (14.8,0.2);
		\draw[->,line width=1.5] (14.5,0.5) to (14.2,0.8);
		\draw[->,line width=1.5] (14.5,0.5) to (14.8,0.8);

		\draw[line width = 1.5] (0,0) to (1,1);
		\draw[line width = 1.5] (1,0) to (0,1);
		\draw[line width = 1.5] (2,0) to (3,1);
		\draw[line width = 1.5] (3,0) to (2,1);
		\draw[line width = 1.5] (4,0) to (5,1);
		\draw[line width = 1.5] (5,0) to (4,1);
		\draw[line width = 1.5] (6,0) to (7,1);
		\draw[line width = 1.5] (7,0) to (6,1);
		\draw[line width = 1.5] (8,0) to (9,1);
		\draw[line width = 1.5] (9,0) to (8,1);
		\draw[line width = 1.5] (10,0) to (11,1);
		\draw[line width = 1.5] (11,0) to (10,1);
		\draw[line width = 1.5] (12,0) to (13,1);
		\draw[line width = 1.5] (13,0) to (12,1);
		\draw[line width = 1.5] (14,0) to (15,1);
		\draw[line width = 1.5] (15,0) to (14,1);
  

  \end{tikzpicture}
	\end{adjustbox}
  \caption{The 8 allowed vertex configurations.}
  \label{figure:vertexxyconfig}
\end{figure}

\begin{figure}[H]
\begin{center}
\includegraphics[width=0.8\textwidth,height=0.45\textheight]{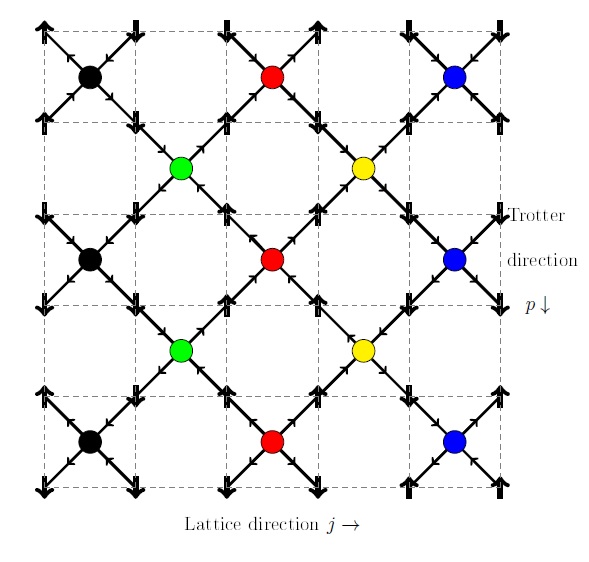}
\end{center}
\caption{Lattice representation demonstrating how configurations of spins on the dotted vertices (represented by arrows $\uparrow \downarrow$) give rise to arrow configurations about the solid vertices.}
 \label{figure:vertexxy}
\end{figure}

As such, the partition function can be thought of as corresponding to a class of two-dimensional classical vertex models on a $\left( \frac{M}{2} + n \right) \times \left( \frac{M}{2} + n\right)$ lattice as shown in Figure~\ref{figure:vertexxy}, with $Mn$ vertices each with a weight \\
$\omega^{j} \left( s_{j,p} s_{j+1,p}, s_{j,p+1}, s_{j+1,p+1} \right)$ given by one of the following
\begin{equation}
\begin{aligned}
&\omega^{j}_{1}(+1,+1,+1,+1) = e^{\frac{h \beta_{\mbox{\tiny qu}}}{n}} \cosh{\left( \frac{2 \beta_{\mbox{\tiny qu}} \gamma}{n} B_{j}\right)}, \\
&\omega^{j}_{2}(-1,-1,-1,-1,) = e^{-\frac{\beta_{\mbox{\tiny qu}} h}{n}} \cosh{\left(\frac{2 \gamma \beta_{\mbox{\tiny qu}}}{n} B_{j}\right)}, \\
&\omega^{j}_{3}(-1,+1,+1,-1) = \omega^{j}_{4}(+1,-1,-1,+1) = \sinh{\left(\frac{ 2 \beta_{\mbox{\tiny qu}}}{n} A_{j}\right)}, \\
& \omega^{j}_{5}(+1,-1,+1,-1) = \omega^{j}_{6}(-1,+1,-1,+1) =  \cosh{\left(\frac{2 \beta_{\mbox{\tiny qu}}}{n} A_{j}\right)}, \\
& \omega^{j}_{7}(-1,-1,+1,+1) = \omega^{j}_{8} (+1,+1,-1,-1) = \sinh{\left( \frac{2 \beta_{\mbox{\tiny qu}} \gamma}{n} B_{j}\right)}, 
\label{vwi}
\end{aligned}
\end{equation}
thus leading to a class of 8-vertex models with the usual 8 possible respective vertex configurations as shown in Figure~\ref{figure:vertexxyconfig}.

The values of these weights depend upon the column $j=1, \ldots, M$ of the original lattice, thus each column has its own separate set of $8$ weights, as represented by the different colours of the circles at the vertices in each column in Figure~\ref{figure:vertexxy}.

Once again, this mapping holds in the limit $n \rightarrow \infty$, which would result in weights $\omega^{i}_{3}, \omega^{i}_{4}, \omega^{i}_{7}, \omega^{i}_{8} \rightarrow 0$ and weights $\omega^{i}_{1}, \omega^{i}_{2}, \omega^{i}_{5}, \omega^{i}_{6} \rightarrow 1$ unless we also take $\beta_{\mbox{\tiny qu}} \rightarrow \infty$. It thus gives us a connection between the ground state properties of the class of quantum systems and the finite temperature properties of the corresponding classical systems. 

\subsection{Algebraic form for the classical partition function}

Finally one last form for the partition function can be obtained using the same method as in Section~\ref{sec:conds} such that the quantum partition function is mapped to one involving entries from matrices given by \eqref{xymatrix}. This time however, instead of applying the extra constraint \eqref{cond}, we can write the partition function as
\begin{equation}
\begin{aligned}
Z = & \lim_{n \rightarrow \infty} \sum_{\sigma_{j,p}=\pm1} \frac{1}{4} \left(\prod^{n}_{p \in a} \prod^{M}_{j \in a} + \prod^{n}_{p \in b} \prod^{M}_{j \in b} \right) \\
& \qquad [ \left(1-s_{j,p} s_{j+1,p} \right) \left( 1+ s_{j,p} s_{j,p+1} \right) \cosh{\frac{2 \beta_{\mbox{\tiny qu}}}{n} A_{j,j+1}}\\
& \qquad + \left(1-s_{j,p} s_{j+1,p} \right) \left(1-s_{j,p} s_{j,p+1} \right) \sinh{\frac{2 \beta_{\mbox{\tiny qu}}}{n} A_{j,j+1}} \\
& \qquad + \left(1+ s_{j,p} s_{j+1,p} \right) \left(1- s_{j,p} s_{j,p+1}\right) \sinh{\frac{ 2 \beta_{\mbox{\tiny qu}} \gamma}{n} B_{j,j+1}} \\
& \qquad + \left(1+ s_{j,p} s_{j,p+1} \right) \left( 1+ s_{j,p} s_{j,p+1} \right) e^{\frac{\beta_{\mbox{\tiny qu}}}{n} h s_{j,p}} \cosh{\frac{ 2 \beta_{\mbox{\tiny qu}} \gamma}{n} B_{j,j+1}} ].
\end{aligned}
\end{equation}

\subsection{Longer range interactions}

The Trotter-Suzuki mapping can similarly be applied to the class of quantum systems \eqref{generalh} with longer range interactions, to obtain partition functions equivalent to classical systems with rather cumbersome descriptions, examples of which can be found in Appendix~\ref{sec:lri}.

\section{Method of coherent states}
\label{sec:coh}

To use the method of coherent states for spin operators $\hat{S}^{i} =\frac{\hbar}{2} \sigma^{i}$, we first apply the Jordan-Wigner transformations \eqref{mbtrans} once more to map the Hamiltonian \eqref{generalh} onto one involving Pauli operators $\sigma^{i}$, $i \in {x,y,z}$;
\begin{equation}
\begin{aligned}
\mathcal{H} = \frac{1}{2} \sum_{1 \leq j \leq k \leq M}   ( & \left(A_{j,k} +\gamma B_{j,k} \right) \sigma^{x}_{j}  \sigma^{x}_{k} \\
& + \left(A_{j,k} - \gamma B_{j,k} \right) \sigma^{y}_{j}  \sigma^{y}_{k} ) \left( \prod^{k-1}_{l=j+1} -\sigma^{z}_{l} \right)  - h\sum^{M}_{j=1}\sigma^{z}_{j}.
\label{generalhpauli}
\end{aligned}
\end{equation}

We then construct a path integral expression for the quantum partition function for \eqref{generalhpauli}. First we divide the quantum
partition function into $n$ pieces
\begin{equation}
Z = \mbox{Tr } e^{-\beta \hat{\mathcal{H}}_{\mbox{\tiny qu}}} = \mbox{Tr } \left[ e^{-\Delta \tau \hat{\mathcal{H}}_{\mbox{\tiny qu}}}  e^{-\Delta \tau \hat{\mathcal{H}}_{\mbox{\tiny qu}}} \ldots  e^{-\Delta \tau \hat{\mathcal{H}}_{\mbox{\tiny qu}}} \right] = \mathbf{V}^{n},
\label{pifz}
\end{equation}
where $\Delta \tau = \frac{\beta}{n}$ and $\mathbf{V}= e^{-\Delta \tau \hat{\mathcal{H}}_{\mbox{\tiny qu}}}$.
 
Next we insert resolutions of the identity in the infinite set of spin coherent states $\ket{\mathbf{N}}$ between each of the $n$ factors in \eqref{pifz}. The coherent states for spin operators, labeled by the continuous vector $\mathbf{N}$ in three-dimensions can be visualised as a classical spin (unit vector) pointing in direction $\mathbf{N}$ such that they have the property
\begin{equation}
\bra{\mathbf{N}} \hat{\mathbf{S}} \ket{\mathbf{N}} = \mathbf{N}.
\end{equation}
They are constructed by applying a rotation operator to an initial state to obtain all the other states as described in \cite{sachdev} such that we end up with
\begin{equation}
\bra{\mathbf{N}} \hat{S}^{i} \ket{\mathbf{N}} = - S N^{i},
\end{equation}
with $N^{i}$s given by 
\begin{equation}
\begin{aligned}
\mathbf{N} & = (N^{x}, N^{y}, N^{z}),  \qquad 0 \leq \theta \leq \pi, \quad 0 \leq \phi \leq 2 \pi, \\
& = (\sin \theta \cos \phi, \sin \theta \sin \phi, \cos \theta).
\label{nparam}
\end{aligned}
\end{equation}

Inserting these states between the $n$ factors in \eqref{pifz} and taking the limit $n \rightarrow \infty$ we obtain
\begin{equation}
Z = \int^{\mathbf{N}(\beta)}_{\mathbf{N}(0)} \mathcal{D} \mathbf{N}(\tau) e^{- \int^{\beta}_{0} d \tau \mathcal{H}(\mathbf{N}(\tau)) - \mathcal{S}_{B}}, 
\label{zfunc}
\end{equation}
where $\mathcal{H}\left( \mathbf{N} \left( \tau \right) \right)$ now has the form of a Hamiltonian corresponding to a two-dimensional classical system given by
\begin{equation}
\begin{aligned}
& \mathcal{H} \left( \mathbf{N} \left( \tau \right) \right) = \bra{\mathbf{N}\left( \tau \right )} \hat{\mathcal{H}}_{\mbox{\tiny qu}} \ket{\mathbf{N} \left( \tau \right)} \\
 = & \sum_{1 \leq j \leq k \leq M} ( \left(A_{j,k} +\gamma B_{j,k} \right) N^{x}_{j}\left( \tau \right)   N^{x}_{k}\left( \tau \right)  \\
& \quad + \left(A_{j,k} - \gamma B_{j,k} \right) N^{y}_{j}\left( \tau \right)   N^{y}_{k} \left( \tau \right) ) \prod^{k-1}_{l=j+1} \left( - N^{z}_{l}\left( \tau \right)  \right) - h \sum^{M}_{j=1} N^{z}_{j}\left( \tau \right) ,\\
= & \sum_{1 \leq j \leq k \leq M}  \left( A_{j,k} \cos \left( \phi_{j}\left( \tau \right)- \phi_{k}\left( \tau \right) \right) + B_{j,k} \gamma \cos \left(\phi_{j}\left( \tau \right)  + \phi_{k}\left( \tau \right) \right) \right) \\ 
& \times \sin \left( \theta_{j} \left( \tau \right) \right) \sin \left( \theta_{k}\left( \tau \right) \right)\prod^{k-1}_{l=j+1} \left( - \cos \left( \theta_{l}\left( \tau \right) \right)\right)  - h \sum^{M}_{j=1} \cos \left( \theta_{j}\left( \tau \right) \right).
\end{aligned}
\end{equation}
The term
\begin{equation}
\mathcal{S}_{B} = \int^{\beta}_{0} d \tau \bra{\mathbf{N}(\tau)} \frac{d}{d \tau} \ket{\mathbf{N}(\tau)}
\label{bph}
\end{equation}
appears through the overlap between the coherent states at two infinitesimally separated steps $\Delta \tau = \tau_{i+1} - \tau_{i}$, and is purely imaginary. This is the appearance of the Berry phase in the action~ \cite{sachdev,wen}. Despite being imaginary, this term gives the correct equation of motion for spin systems \cite{wen}. 


\section{Simultaneous diagonalisation of the quantum Hamiltonian and the transfer matrix}
\label{sec:simdiag}

This section presents a particular type of equivalence between one-dimensional quantum and two-dimensional classical models, established by commuting the quantum Hamiltonian with the transfer matrix of the classical system under certain parameter relations between the corresponding systems. Suzuki \cite{xyi} used this method to prove an equivalence between the one-dimensional generalised quantum XY model and the two-dimensional Ising and dimer models under specific parameter restrictions between the two systems. In particular he proved that this equivalence holds when the quantum system is restricted to nearest neighbour or nearest and next nearest neighbour interactions.

Here we extend the work of Suzuki \cite{xyi}, establishing this type of equivalence between the class of quantum spin chains \eqref{generalh} for all interaction lengths when restricted to $U(N)$ symmetry\footnote{which we see from Table~\ref{tab:ccgg} means the matrices $A_{j,k}$ and $B_{j,k}$ have Toeplitz structure.}, and the two-dimensional Ising and dimer models under certain restrictions amongst coupling parameters. For the Ising model we use both transfer matrices forming two separate sets of parameter relations under which the systems are equivalent. Where possible we connect critical properties of the corresponding systems, providing a pathway with which to show that the critical properties of these classical systems are also influenced by symmetry.

All discussions regarding the general class of quantum systems \eqref{generalh} in this section refer to the family corresponding to $U(N)$ symmetry only, in which case, we find that
\begin{equation}
\left[\mathcal{H}_{\mbox{\tiny qu}},\mathbf{V}_{\mbox{\tiny cl}}\right] = 0,
\label{commtrans}
\end{equation}
under appropriate relationships amongst parameters of the quantum and classical systems, when $\mathbf{V}_{\mbox{\tiny cl}}$ is the transfer matrix for either the two-dimensional Ising model with Hamiltonian given by
\begin{equation}
\mathcal{H} = - \sum^{N}_{i} \sum^{M}_{j} \left( J_{1} s_{i,j} s_{i+1,j} + J_{2} s_{i,j} s_{i,j+1} \right),
\label{2disingh}
\end{equation}
or the dimer model. 

A dimer is a rigid rod covering exactly two neighbouring vertices either vertically or horizontally. The model we refer to is one consisting of a square planar lattice with $N$ rows and $M$ columns, with an allowed configuration being when each of the $N M$ vertices is covered exactly once such that 
\begin{equation}
2h + 2v = NM,
\label{dimsite}
\end{equation}
where $h$ and $v$ are the number of horizontal and vertical dimers respectively. The partition function is given by
\begin{equation}
\begin{aligned}
Z & = \sum_{\mbox{\tiny allowed configs}} x^{h} y^{v}\\
& = y^{\frac{MN}{2}} \sum_{\mbox{\tiny allowed configs}} \alpha^{h},
\end{aligned}
\end{equation}
where $x$ and $y$ are the appropriate ``activities'' and $\alpha = \frac{x}{y}$.

The transform used to diagonalise both of these classical systems as well as the class of quantum spin chains \eqref{generalh} can be written as
\begin{equation}
\begin{aligned}
\eta^{\dagger}_{q} & = \frac{e^{-\frac{i \pi}{4}} }{\sqrt{M}} \sum_{j} e^{-\frac{2 \pi i}{M}qj} \left(b^{\dagger}_{j} u_{q} + i b_{j} v_{q} \right),\\
\eta_{q} & = \frac{e^{\frac{i \pi}{4}} }{\sqrt{M}} \sum_{j} e^{\frac{2 \pi i}{M}qj} \left(b_{j} u_{q} -i b^{\dagger}_{j} v_{q} \right),
\end{aligned}
\label{lintransis}
\end{equation}
where the $\eta_{q}$s are the Fermi operators in which the systems are left in diagonal form. This diagonal form is given by Eq.\eqref{hdiag} for the quantum system and for the transfer matrix for the Ising model by\footnote{The superscripts $^{+(-)}$ represent anticyclic and cyclic boundary conditions respectively.} \cite{2dising} 
\begin{equation}
\mathbf{V}^{+(-)} = \left(2 \sinh 2 K_{1} \right)^{\frac{N}{2}}  e^{-\sum_{q} \epsilon_{q} \left(\eta^{\dagger}_{q} \eta_{q} - \frac{1}{2}\right)},
\label{diagform2dclis}
\end{equation}
where $K_{i}=\beta J_{i}$ and $\epsilon_{q}$ is the positive root of\footnote{This is for the symmetrisation $\mathbf{V}=\mathbf{V}^{\frac{1}{2}}_{1} \mathbf{V}_{2} \mathbf{V}^{\frac{1}{2}}_{1}$ of the transfer matrix; the other possibility is with $\mathbf{V}'=\mathbf{V}^{\frac{1}{2}}_{2} \mathbf{V}_{1} \mathbf{V}^{\frac{1}{2}}_{2}$ where $\mathbf{V}_{1} = \left(2 \sinh 2 K_{1} \right)^{\frac{M}{2}} e^{-K^{*}_{1} \sum^{M}_{i} \sigma^{x}_{i}}$, $V_{2} = e^{K_{2} \sum^{M}_{i=1} \sigma^{z}_{i} \sigma^{z}_{i+1}}$, and $\tanh{K^{*}_{i}} = e^{-2K_{i}}$.}
\begin{equation}
\cosh \epsilon_{q} = \cosh 2K^{*}_{1} \cosh 2 K_{2} - \sinh 2K^{*}_{1} \sinh 2 K_{2} \cos q. 
\end{equation}

The dimer model on a two-dimensional lattice was first solved by Kasteleyn \cite{dimerk} via a combinatorial method reducing the problem to the evaluation of a Pfaffian. Lieb \cite{dimerl} later formulated the dimer-monomer problem in terms of transfer matrices such that $\mathbf{V}_{\mbox{\tiny cl}} =\mathbf{V}^{2}_{\mbox{\tiny D}}$ is left in the diagonal form given by
\begin{equation}
\mathbf{V}^{2}_{D} = \prod_{0 \leq q \leq \pi} \left( \lambda_{q} \left( \eta^{\dagger}_{q} \eta_{q} + \eta^{\dagger}_{-q} \eta_{-q} -1 \right) + \left( 1 + 2 \alpha^{2}\sin^{2}q\right) \right),
\label{tmdim}
\end{equation}
with
\begin{equation}
\lambda_{q} =2 \alpha \sin q \left(1 + \alpha^{2} \sin^{2} q \right)^{\frac{1}{2}}.
\end{equation}

For the class of quantum spin chains \eqref{generalh} as well as each of these classical models we have that the ratio of terms in the transform \eqref{lintransis} is given by
\begin{equation}
\frac{2v_{q}u_{q}}{u^{2}_{q}-v^{2}_{q}} =
\begin{cases}
\frac{a_{q}}{b_{q}} & \quad \mbox{for } \mathcal{H}_{\mbox{\tiny qu}} \\
\frac{\sin q}{\cosh 2K^{*}_{1}\cos q - \sinh 2K^{*}_{1} \coth 2K_{2} } & \quad \mbox{for } \mathbf{V}, \\
\frac{\sin q \left( 1 - \tanh 2 K^{*}_{1} \tanh K_{2} \cos q \right)}{\cos q - \tanh K_{2} \tanh 2 K^{*}_{1} \cos^{2}q - \frac{\tanh 2 K^{*}_{1} }{\sinh 2 K_{2}}} & \quad \mbox{for } \mathbf{V'},\\
-\frac{1}{\alpha \sin q} & \quad \mbox{for } \mathbf{V}_{D}^{2},
\end{cases}
\label{tanis}
\end{equation}
which we show in the following sections will provide us with relationships between parameters under which these classical systems are equivalent to the quantum systems. 

\subsection{The Ising model with transfer matrix $\mathbf{V}$}

We see from \eqref{tanis} that the Hamiltonian \eqref{generalh} commutes with the transfer matrix $\mathbf{V}$ if we require that
\begin{equation}
\frac{a_{q}}{b_{q}} = \frac{\sin q}{\cosh 2K^{*}_{1}\cos q - \sinh 2K^{*}_{1} \coth 2K_{2} }.
\end{equation}

This provides us with the following relations between parameters under which this equivalence holds\footnote{Here we have used the De Moivre's Theorem and the binomial formula to rewrite the summations in $a_{q}$ and $b_{q}$ \eqref{aborig} as
\begin{equation}
\begin{aligned}
a_{q} &=  \Gamma + \sum^{L}_{k=1} a(k) \sum^{\left[\frac{k}{2}\right]}_{l=0} \binom{k}{2l} \sum^{l}_{i=0} \binom{l}{i} \left(-1\right)^{-i} \cos^{k-2i}q,\\
b_{q} &= \tan q \sum^{L}_{k=1} b(k) \sum^{\left[\frac{k-1}{2}\right]}_{l=0} \binom{k}{2l+1} \sum^{l}_{i=0} \binom{l}{i}\left(-1\right)^{-i} \cos^{k-2i} q.
\end{aligned}
\end{equation}}
\begin{equation}
\begin{aligned}
& \sinh 2K^{*}_{1} \coth 2K_{2} = -\frac{\bar{a}(L-1)}{ \bar{b}(L)}, \quad \tanh^{2} K^{*}_{1} = \frac{ \bar{a}(L) - \bar{b}(L)}{ \bar{a}(L) + \bar{b}(L) } \\
& \quad \mbox{and} \quad \frac{ \bar{a}(L-1)}{ \bar{a}(L) + \bar{b}(L)} = - \coth 2K_{2}\tanh K^{*}_{1},
\label{paramrel}
\end{aligned}
\end{equation}
or inversely as
\begin{equation}
\cosh 2K^{*}_{1} = \frac{\bar{a}(L)}{ \bar{b}(L)}, \quad \tanh 2K_{2} = - \frac{1 }{\bar{a}(L-1)} \sqrt{\left(\bar{a}(L)\right)^{2} - \left( \bar{b}(L)\right)^{2}},
\label{paramrelin}
\end{equation}
where
\begin{equation}
\bar{a}(L) = a(L) \sum^{\left[\frac{L}{2} \right]}_{l=0} \binom{L}{2l}, \quad \bar{b}(L) =b(L) \sum^{\left[\frac{L-1}{2} \right]}_{l=0} \binom{L}{2l+1},
\label{abmm}
\end{equation}
and 
\begin{equation}
\bar{a}(0) = \Gamma.
\label{a00}
\end{equation}

From \eqref{paramrelin} we see that this equivalence holds when
\begin{equation}
\frac{\bar{a}(L)}{\bar{b}(L)} \geq 1, \qquad \bar{a}^{2}(L)  \leq \bar{a}^{2}(L-1) + \bar{b}^{2}(L).
\end{equation}

For $L>1$, we also have the added restrictions on the parameters that
\begin{equation}
\begin{aligned}
\sum^{L}_{k=1} b(k) \sum^{\left[\frac{L-1}{2}\right]}_{l=0} \binom{k}{2l+1} \sum^{l}_{i=1} (-1)^{i} \cos^{k-2i}q + \sum^{L-1}_{k=1} \bar{b}(k) \cos^{k}q =0,
\label{bcoeffs}
\end{aligned}
\end{equation}
and
\begin{equation}
\begin{aligned}
\Gamma + \sum^{L-2}_{k=1} \bar{a}(k) \cos^{k}q + \sum^{L}_{k=1} a(k) \sum^{\left[\frac{k}{2}\right]}_{l=0} \binom{k}{2l} \sum^{l}_{i=1} \binom{l}{i} \left(-1\right)^{i} \cos^{L-2i}q =0,
\label{acoeffs}
\end{aligned}
\end{equation} 
which implies that all coefficients of $\cos^{i}q$  for $0 \leq i<L$ in \eqref{bcoeffs} and of $\cos^{i}q$ for $0 \leq i<L-1$ in \eqref{acoeffs} are zero\footnote{For example setting the coefficient of $\left(\cos q \right)^{0}$ to zero implies that $\Gamma = - \sum^{\left[\frac{L-1}{2}\right]}_{j=1} (-1)^{j} a(2j)$.}.

When only nearest neighbour interactions are present in \eqref{generalh} ($L=1$), with $a(k)=b(k) = 0$ for $k\neq 1$ we recover Suzuki's result~\cite{xyi}.

The critical properties of the class of quantum systems can be analysed from the dispersion relation~\eqref{lambda}, which under the above parameter restrictions is given by 
\begin{equation}
\left| \Lambda_{q} \right| = 2^{p+1} \left| \cos^{(L-1)} q \right| \left( \left( \bar{a}(L) \cos q + \bar{a}(L-1) \right)^{2} + \bar{b}^{2}(L) \sin^{2} q \right)^{\frac{1}{2}},
\label{lq3}
\end{equation}
which is gapless for $L>1$ for all parameter values.

The critical temperature for the Ising model~\cite{2dising} is given by
\begin{equation}
K^{*}_{1} = K_{2},
\label{k1k2}
\end{equation}
which using \eqref{paramrel} and \eqref{paramrelin} gives
\begin{equation}
\begin{aligned}
\bar{a}(L) & = \pm \bar{a}(L-1).
\label{crittempis}
\end{aligned}
\end{equation}

This means that \eqref{lq3} becomes
\begin{equation}
\left| \Lambda_{q} \right| = 2^{p+1} \left| \bar{a}(L) \cos^{(L-1)} q \right| \left( \left( \cos q \pm 1 \right)^{2} + \left(\frac{\bar{b}(L)}{\bar{a}(L)}\right)^{2} \sin^{2} q \right)^{\frac{1}{2}},
\end{equation}
which is now gapless for all $L > 1$, and for $L=1$, \eqref{crittempis} is the well known critical value for the external field for the quantum XY model.

The correlation function between two spins in the same row in the classical Ising model at finite temperature can also be written in terms of those in the ground state of the quantum model;
\begin{equation}
\begin{aligned}
\left\langle \sigma^{x}_{j,k} \sigma^{x}_{j+r,k} \right\rangle_{\mbox{\tiny Is}} & = \bra{\Psi_{0}}  \mathbf{V}_{1}^{-\frac{1}{2}} \sigma^{x}_{j} \sigma^{x}_{j+r} \mathbf{V}_{1}^{\frac{1}{2}} \ket{\Psi_{0}} \\
& = \bra{\Phi_{0}} \mathbf{V}_{1}^{-\frac{1}{2}} \sigma^{x}_{j} \sigma^{x}_{j+r} \mathbf{V}_{1}^{\frac{1}{2}} \ket{\Phi_{0}}\\
& = \left\langle \left( \mathbf{V}_{1}^{-\frac{1}{2}}  \sigma^{x}_{j} \mathbf{V}_{1}^{\frac{1}{2}}  \right) \left( \mathbf{V}_{1}^{-\frac{1}{2}}  \sigma^{x}_{j+r} \mathbf{V}_{1}^{\frac{1}{2}}  \right) \right\rangle_{\mbox{\tiny qu}} \\
&= \cosh^{2} K^{*}_{1} \left\langle \sigma^{x}_{j} \sigma^{x}_{j+r} \right\rangle_{\mbox{\tiny qu}} - \sinh^{2} K^{*}_{1} \left\langle \sigma^{y}_{j} \sigma^{y}_{j+r} \right\rangle_{\mbox{\tiny qu}},
\end{aligned}
\end{equation}
using the fact that $\left\langle \sigma^{x}_{j} \sigma^{y}_{j+r} \right\rangle_{\mbox{\tiny qu}} = \left\langle \sigma^{y}_{j} \sigma^{x}_{j+r} \right\rangle_{\mbox{\tiny qu}} = 0$, for $r \neq 0$, and
\begin{equation}
\Psi_{0} = \Phi_{0},
\label{psiphi}
\end{equation}
from~\eqref{hdiag}, \eqref{commtrans} and \eqref{diagform2dclis}, where $\Psi_{0}$ is the eigenvector corresponding to the maximum eigenvalue of $\mathbf{V}$ and $\Phi_{0}$ is the ground state eigenvector for the general class of quantum systems \eqref{generalh} (restricted to $U(N)$ symmetry). 

This implies that the correspondence between critical properties (i.e. correlation functions), is not limited to quantum systems with short range interactions (as Suzuki~\cite{xyi} found), but also holds for a more general class of quantum systems for a fixed relationship between the magnetic field and coupling parameters as dictated by \eqref{acoeffs} and \eqref{bcoeffs}, which we see from \eqref{lq3} results in a gapless system. 

\subsection{The Ising model with transfer matrix $\mathbf{V}'$}

From \eqref{tanis} the Hamiltonian for the quantum spin chains \eqref{generalh} commutes with transfer matrix $\mathbf{V'}$ if we set
\begin{equation}
\frac{a_{q}}{b_{q}} = \frac{\sin q \left( 1 - \tanh 2 K^{*}_{1} \tanh K_{2} \cos q \right)}{\cos q - \tanh K_{2} \tanh 2 K^{*}_{1} \cos^{2}q - \frac{\tanh 2 K^{*}_{1} }{\sinh 2 K_{2}}}. 
\end{equation}

This provides us with the following relations between parameters under which this equivalence holds when the class of quantum spin chains \eqref{generalh} has an interaction length $L>1$;
\begin{equation}
\begin{aligned}
& \tanh 2 K^{*}_{1} \tanh K_{2} = -\frac{\bar{b}(L)}{\bar{b}(L-1)} = - \frac{\bar{a}(L)}{\bar{b}(L-1)} \\
&\mbox{and} \quad \frac{\bar{a}(L-1)}{\bar{b}(L-1)} =1, \quad \frac{\tanh 2K^{*}_{1}}{\sinh 2 K_{2}} = - \frac{\bar{a}^{*}(L)}{\bar{b}(L-1)},
\label{paramrelvnd}
\end{aligned}
\end{equation}
or inversely as
\begin{equation}
\begin{aligned}
\sinh^{2} K_{2} & = \frac{\bar{a}(L)}{2\left(\bar{a}^{*}(L)\right)} \quad \mbox{and}\\
\tanh2K^{*}_{1} &= - \frac{1}{\bar{a}(L-1)}\sqrt{\bar{a}(L)\left(2\bar{a}^{*}(L)+a(L)\right)},
\label{paramrelex}
\end{aligned}
\end{equation}
where
\begin{equation}
\bar{a}^{*}(L) = \bar{a}(L-2) -a(L) \sum^{\left[\frac{L}{2} \right]}_{l=0} \binom{L}{2l} l.
\end{equation}

From \eqref{paramrelex} we see that this equivalence holds when
\begin{equation}
\bar{a}(L)\left(2\bar{a}^{*}(L)+\bar{a}(L)\right) \leq \bar{a}^{2}(L-1).
\end{equation}

When $L>2$, we have further restrictions upon the parameters of the class of quantum systems \eqref{generalh}, namely
\begin{equation}
\sum^{L-2}_{k=1} \bar{b}(k) \cos^{k}q + \sum^{L}_{k=1} b(k) \sum^{\left[\frac{k-1}{2}\right]}_{l=0} \binom{k}{2l+1} \sum^{l}_{i=1} \binom{l}{i} (-1)^{i} \cos^{k-2i}q =0
\label{bbcoeffs}
\end{equation}
and
\begin{equation}
\begin{aligned}
& \Gamma + \sum^{L-3}_{k=1} \bar{k} \cos^{k}q - \sum^{L-1}_{k=1} a (k) \sum^{\left[\frac{k}{2}\right]}_{l=0} \binom{k}{2l} l \cos^{k-2} q \\
& +\sum^{L}_{k=1} a(k) \sum^{\left[\frac{k}{2}\right]}_{l=0} \binom{k}{2l} \sum^{l}_{i=2} \binom{l}{i}(-1)^{i} \cos^{k-2i}q =0.
\label{aacoeffs}
\end{aligned}
\end{equation}
This implies that coefficients of $\cos^{i}q$ for $0 \leq i<L-1$ in \eqref{bbcoeffs} and of $\cos^{i}q$ for $0 \leq i<L-2$ in \eqref{aacoeffs} are zero.

Under these parameter restrictions, the dispersion relation is given by
\begin{equation}
\begin{aligned}
\left| \Lambda_{q} \right|  = 2^{p+1} \left| \cos^{L-2} q \right| & ( \left( \cos q \left( \bar{a}(L) \cos q + \bar{a}(L-1) \right) + \bar{a}^{*}(L) \right)^{2} \\
& + \sin^{2} q \left(\bar{b}(L) \cos q + \bar{b}(L-1) \right) )^{\frac{1}{2}},
\label{lq6}
\end{aligned}
\end{equation}
which is gapless for $L>2$ for all parameter values.

The critical temperature for the Ising model \eqref{k1k2} becomes
\begin{equation}
-\bar{a}(L-1) = \bar{a}^{*}(L)+ \bar{a}(L),
\label{aaa}
\end{equation}
using \eqref{paramrelvnd} and \eqref{paramrelex}.

Substituting \eqref{aaa} into \eqref{lq6} we obtain
\begin{equation}
\begin{aligned}
\left| \Lambda_{q} \right|  = 2^{p+1} \left| \cos^{L-2} q \right| & (\left( \bar{a}(L) \cos q - \bar{a}^{*}(L) \right)^{2} \left( \cos q -1 \right)^{2} \\
& + \sin^{2} q \left(\bar{b}(L) \cos q + \bar{b}(L-1) \right) )^{\frac{1}{2}},
\end{aligned}
\end{equation}
which we see is now gapless for all $L \geq 2$ (for $L=2$ this clearly corresponds to a critical value of $\Gamma$ causing the energy gap to close).

In this case we can once again write the correlation function for spins in the same row of the classical Ising model at finite temperature in terms of those in the ground state of the quantum model as
\begin{equation}
\left\langle \sigma^{x}_{j,k} \sigma^{x}_{j+r,k} \right\rangle_{\mbox{\tiny Is}} = \left\langle \sigma^{x}_{j} \sigma^{x}_{j+r} \right\rangle_{\mbox{\tiny qu}},
\end{equation}
where $\Psi'_{0}$ is the eigenvector corresponding to the maximum eigenvalue of $\mathbf{V'}$ and 
\begin{equation}
\Psi'_{0} = \Phi_{0}.
\end{equation}

Once more this implies that the correspondence between critical properties such as correlation functions is not limited to quantum systems with short range interactions; it also holds for longer range interactions, for a fixed relationship between the magnetic field and coupling parameters which causes the systems to be gapless. 

\subsection{The dimer model with transfer matrix $\mathbf{V}^{2}_{D}$}

In this case, when the class of quantum spin chains \eqref{generalh} has a maximum interaction length $L>1$, it is possible to find relationships between parameters for which an equivalence is obtained between it and the two-dimensional dimer model. For details and examples see Appendix~\ref{sec:appapprelsvd}.

For $a(k)=b(k)=0$ for $k>2$, we recover Suzuki's result~\cite{xyi}.

\section{Acknowledgements}
We are grateful to Professor Shmuel Fishman for helpful discussions. JH is pleased to thank Nick Jones for several insightful remarks, to the EPSRC for support during her PhD, and to the Leverhulme Trust for further support. FM was partially supported by EPSRC research grants EP/G019843/1 and EP/L010305/1.

\begin{appendices}
\section{Symmetry classes}
\label{sec:appsymcl}
\begin{table}[H]
\centering
\begin{tabular}{c c c}
\hline \hline
Classical compact & Structure of matrices & Matrix entries \\
 group & $\bar{A}_{j,k} \quad (\bar{B}_{j,k})$ & $\left( \mathcal{M}_{n} \right)_{j,k} $ \\
\hline \hline
$U(N)$ & $a(j-k) \quad (b(j-k))$ & $g_{j-k}, \quad j,k \geq 0$ \\
$O^{+}(2N)$ & $a(j-k)+a(j+k)$ & $g_{0} \quad$ if $j=k=0$\\
 & & $\sqrt{2} g_{l}$  if \\
& &  either $j=0, k=l$ \\
& & $\quad$ or $j=l, k=0$ \\
 & & $g_{j-k} + g_{j+k}, \quad j,k >0$ \\
$Sp(2N)$ & $a(j-k) -a(j+k+2)$ & $g_{j-k}-g_{j+k+2}, \quad j, k \geq 0$ \\
$O^{\pm}(2N+1)$ & $a(j-k) \mp a(j+k+1)$ & $g_{j-k} \mp g_{j+k+1}, \quad j,k \geq 0$\\ 
$O^{-}(2N+2)$ & $a(j-k) -a(j+k+2)$ & $g_{j-k} - g_{j+k+2}, \quad j,k \geq 0 $ \\
\hline
\end{tabular}
\caption{The structure of functions $a(j)$ and $b(j)$ dictating the entries of matrices $\mathbf{\bar{A}} = \mathbf{A}- 2 h \mathbf{I}$ and $\mathbf{\bar{B}} = \gamma \mathbf{B}$, which reflect the respective symmetry groups. The $g_{l}$s are the Fourier coefficients of the symbol $g^{\mathcal{M}} \left( \theta \right)$ of $\mathcal{M}_{M}$. Note that for all symmetry classes other than $U(N)$, $\gamma=0$ and thus $\bar{\mathbf{B}}=0$.}
\label{tab:ccgg}
\end{table}

\section{Longer range interactions}
\label{sec:lri}

\subsubsection{Nearest and next nearest neighbour interactions}

The class of quantum systems \eqref{generalh} with nearest and next nearest neighbour interactions can be mapped\footnote{Once again ignoring boundary term effects due to our interest in phenomena in the thermodynamic limit only.} onto
\begin{equation}
\begin{aligned}
\mathcal{H}_{\mbox{\tiny qu}} = - \sum^{M}_{j=1}( &J^{x}_{j} \sigma^{x}_{j} \sigma^{x}_{j+1} + J^{y}_{j} \sigma^{y}_{j} \sigma^{y}_{j+1} - \left( J'^{x}_{j} \sigma^{x}_{j}\sigma^{x}_{j+2} + J'^{y}_{j} \sigma^{y}_{j} \sigma^{y}_{j+2} \right) \sigma^{z}_{j+1}\\
& + h \sigma^{z}_{j} ),
\label{hnnn}
\end{aligned}
\end{equation}
where $J'^{x}_{j} = \frac{1}{2} \left( A_{j,j+2} + \gamma B_{j,j+2} \right)$ and $J'^{y}_{j} = \frac{1}{2} \left( A_{j,j+2} - \gamma B_{j,j+2} \right)$ using the Jordan Wigner transformations \eqref{mbtrans}.

We apply the Trotter Suzuki mapping to the partition function for \eqref{hnnn} with operators in the Hamiltonian ordered as
\begin{equation}
Z = \lim_{n \rightarrow \infty} \mbox{Tr } \left[ e^{\frac{\beta_{\mbox{\tiny qu}}}{n} \hat{\mathcal{H}}^{x}_{a}} e^{\frac{\beta_{\mbox{\tiny qu}}}{2n} \hat{\mathcal{H}}^{z}} e^{\frac{\beta_{\mbox{\tiny qu}}}{n} \hat{\mathcal{H}}^{y}_{b}} e^{\frac{\beta_{\mbox{\tiny qu}}}{n} \hat{\mathcal{H}}^{y}_{a}} e^{\frac{\beta_{\mbox{\tiny qu}}}{2n} \hat{\mathcal{H}}^{z}} e^{\frac{\beta_{\mbox{\tiny qu}}}{n} \hat{\mathcal{H}}^{x}_{b}} \right]^{n}, 
\label{z1nnn}
\end{equation}
where again $a$ and $b$ are the set of odd and even integers respectively, and \\
$\hat{\mathcal{H}}_{\alpha}^{\mu} = \sum^{M}_{j \in \alpha} \left( \frac{1}{2} J^{\mu}_{j} \left(\sigma^{\mu}_{j} \sigma^{\mu}_{j+1}+  \sigma^{\mu}_{j+1} \sigma^{\mu}_{j+2} \right) - J'^{\mu}_{j} \sigma^{\mu}_{j} \sigma^{z}_{j+1} \sigma^{\mu}_{j+2} \right)$ and $\hat{\mathcal{H}}^{z} = h \sum^{M}_{j=1} \sigma^{z}_{j}$, for $\mu \in x,y$ and once more $\alpha$ denotes either $a$ or $b$.

For this model we need to insert $4n$ identity operators into \eqref{z1nnn}. We use $n$ in each of the $\sigma^{x}$ and $\sigma^{y}$ bases and $2n$ in the $\sigma^{z}$ basis in the following way
\begin{equation}
\begin{aligned}
Z =  \lim_{n \rightarrow \infty} & \mbox{Tr } \left[\mathbb{I}_{\sigma_{1}}  e^{\frac{\beta_{\mbox{\tiny qu}}}{n} \hat{\mathcal{H}}^{x}_{a}} \mathbb{I}_{s_{1}}  e^{\frac{\beta_{\mbox{\tiny qu}}}{2n} \hat{\mathcal{H}}^{z}} e^{\frac{\beta_{\mbox{\tiny qu}}}{n} \hat{\mathcal{H}}^{y}_{b}} \mathbb{I}_{\tau_{1}}  e^{\frac{\beta_{\mbox{\tiny qu}}}{n} \hat{\mathcal{H}}^{y}_{a}} \mathbb{I}_{s_{1}}  e^{\frac{\beta_{\mbox{\tiny qu}}}{2n} \hat{\mathcal{H}}^{z}}  e^{\frac{\beta_{\mbox{\tiny qu}}}{n} \hat{\mathcal{H}}^{x}_{b}} \right]^{n} \\
= \lim_{n \rightarrow \infty} & \sum_{\sigma_{j,p}, \tau_{j,p}, s_{j,p}} \prod^{n}_{p=1} [ \bra{\vec{\sigma}_{p}} e^{\frac{\beta_{\mbox{\tiny qu}}}{n} \hat{\mathcal{H}}^{x}_{a}}  \ket{\vec{s}_{2p-1}} \bra{\vec{s}_{2p-1}} e^{\frac{\beta_{\mbox{\tiny qu}}}{2n} \hat{\mathcal{H}}^{z}} e^{\frac{\beta_{\mbox{\tiny qu}}}{n} \hat{\mathcal{H}}^{y}_{b}} \ket{\vec{\tau}_{p}} \\
& \qquad \bra{\vec{\tau}_{p}} e^{\frac{\beta_{\mbox{\tiny qu}}}{n} \hat{\mathcal{H}}^{y}_{a}} \ket{\vec{s}_{2p}} \bra{\vec{s}_{2p}}  e^{\frac{\beta_{\mbox{\tiny qu}}}{2n}  \hat{\mathcal{H}}^{z}}  e^{\frac{\beta_{\mbox{\tiny qu}}}{n} \hat{\mathcal{H}}^{x}_{b}} \ket{\vec{\sigma}_{p+1}} ].
\label{z2nnn}
\end{aligned}
\end{equation}

For this system it is then possible to rewrite the remaining matrix elements in \eqref{z2nnn} in complex scalar exponential form by first writing 
\begin{equation}
\begin{aligned}
 \bra{\vec{\sigma}_{p}} & e^{\frac{\beta}{n} \hat{\mathcal{H}}^{x}_{a}}  \ket{\vec{s}_{2p-1}} \bra{\vec{s}_{2p-1}}  e^{\frac{\beta_{\mbox{\tiny qu}}}{2n} \hat{\mathcal{H}}^{z}} e^{\frac{\beta_{\mbox{\tiny qu}}}{n} \hat{\mathcal{H}}^{y}_{b}} \ket{\vec{\tau}_{p}} \\
&  \bra{\vec{\tau}_{2p-1}}  e^{\frac{\beta_{\mbox{\tiny qu}}}{n} \hat{\mathcal{H}}^{y}_{a}} \ket{\vec{s}_{2p}} \bra{\vec{s}_{2p}}  e^{\frac{\beta_{\mbox{\tiny qu}}}{2n}\hat{\mathcal{H}}^{z}} e^{\frac{\beta_{\mbox{\tiny qu}}}{n} \hat{\mathcal{H}}^{x}_{b}} \ket{\vec{\sigma}_{2p}} \\
& \quad  =  e^{\frac{\beta_{\mbox{\tiny qu}}}{n} \mathcal{H}^{x}_{a}(p)} e^{\frac{\beta_{\mbox{\tiny qu}}}{2n} \mathcal{H}^{z}(2p-1) }  e^{\frac{\beta_{\mbox{\tiny qu}}}{n} \mathcal{H}^{y}_{b}(p)}  e^{\frac{\beta_{\mbox{\tiny qu}}}{n} \mathcal{H}^{y}_{a}(p)}  e^{\frac{\beta_{\mbox{\tiny qu}}}{2n} \mathcal{H}^{z}(2p)}  e^{\frac{\beta_{\mbox{\tiny qu}}}{n} \mathcal{H}^{x}_{b}(p)}\\
& \qquad \braket{\vec{\sigma}_{p}}{\vec{s}_{2p-1}} \braket{\vec{s}_{2p-1}}{\vec{\tau}_{p}}  \braket{\vec{\tau}_{p}}{\vec{s}_{2p}} \braket{\vec{s}_{2p}}{\vec{\sigma}_{p+1}},
\end{aligned}
\end{equation}
where $\mathcal{H}^{x}_{\alpha}(p) = \sum^{M}_{j \in \alpha} (\frac{1}{2} J^{x}_{j} \left(\sigma_{j,p} \sigma_{j+1,p}+ \sigma_{j+1,p} \sigma_{j+2,p} \right)+ J'^{x}_{j+1} \sigma_{j,p} s_{j+1,p} \sigma_{j+2,p}  ) $, $\mathcal{H}^{y}_{\alpha}(p) = \sum^{M}_{j \in \alpha} (\frac{1}{2} J^{y}_{j} \left(\tau_{j,p} \tau_{j+1,p}+ \tau_{j+1,p} \tau_{j+2,p} \right)+ J'^{y}_{j+1} \tau_{j,p} s_{j+1,p} \tau_{j+2,p}  ) $  and $\mathcal{H}^{z}(p) = \sum^{M}_{j =1}  s_{j,p}$. We can then evaluate the remaining matrix elements as
\begin{equation}
\begin{aligned}
& \braket{\vec{\sigma}_{p}}{\vec{s}_{2p-1}}  \braket{\vec{s}_{2p-1}}{\vec{\tau}_{p}}  \braket{\vec{\tau}_{2p-1}}{\vec{s}_{2p}} \braket{\vec{s}_{2p}}{\vec{\sigma}_{p+1}} \\
& = \frac{1}{2^{4M}} \prod^{M}_{j=1} e^{\frac{i\pi}{4} \left(  -s_{j,2p-1} + s_{j,2p} + \sigma_{j,p} s_{j,2p-1} - \sigma_{j,p+1}s_{2p} + \tau_{j,p} \left( s_{j,2p} - s_{j,2p-1} \right)\right) }.
\end{aligned}
\end{equation}

Thus we obtain a partition function with the same form as that corresponding to a class of two-dimensional classical Ising type systems on a $M \times 4n$ lattice with classical Hamiltonian $\mathcal{H}_{\mbox{\tiny cl}}$ given by
\begin{equation}
\begin{aligned}
& -\beta_{\mbox{\tiny cl}} \mathcal{H}_{\mbox{\tiny cl}} \\
& = \frac{\beta_{\mbox{\tiny qu}}}{n} \sum^{n}_{p =1} (  \sum_{j \in a} \left(\frac{ J^{x}_{j}}{2} \left(\sigma_{j,p} \sigma_{j+1,p} + \sigma_{j+1,p} \sigma_{j+2,p} \right) - J'^{x}_{j+1} \sigma_{j,p} s_{j+1,p} \sigma_{j+2,p}  \right) \\
& + \sum_{j \in b} \left(\frac{J^{y}_{j}}{2} \left(\tau_{j,p} \tau_{j+1,p} + \tau_{j+1,p} \tau_{j+2,p} \right) - J'^{y}_{j+1} \tau_{j,p} s_{j+1,2p-1} \tau_{j+2,p}  \right) \\
& + \sum_{j \in a} \left(\frac{J^{y}_{j}}{2} \left(\tau_{j,p} \tau_{j+1,p} + \tau_{j+1,p} \tau_{j+2,p} \right) - J'^{y}_{j+1} \tau_{j,p} s_{j+1,2p} \tau_{j+2,p}  \right) \\
& + \sum_{j \in b} (\frac{J^{x}_{j}}{2} \left(\sigma_{j,p+1} \sigma_{j+1,p+1} + \sigma_{j+1,p+1} \sigma_{j+2,p+1} \right) \\
& \quad - J'^{x}_{j+1} \sigma_{j,p+1} s_{j+1,2p} \sigma_{j+2,p+1} )) \\
& + \sum^{n}_{p =1} (\sum^{M}_{j=1} \left( \left( \frac{\beta_{\mbox{\tiny qu}} h}{2n} - \frac{i \pi}{4} \right) s_{j,2p} + \left(\frac{\beta_{\mbox{\tiny qu}} h}{2n} +\frac{i \pi}{4} \right) s_{j,2p} \right)\\
& + \sum^{M}_{j=1} \frac{i\pi}{4} \left( \sigma_{j,p} s_{j,2p-1} - \sigma_{j,p+1}s_{2p} + \tau_{j,p} \left( s_{j,2p} - s_{j,2p-1} \right)\right) ) + 4nM \ln 2.
\end{aligned}
\end{equation}

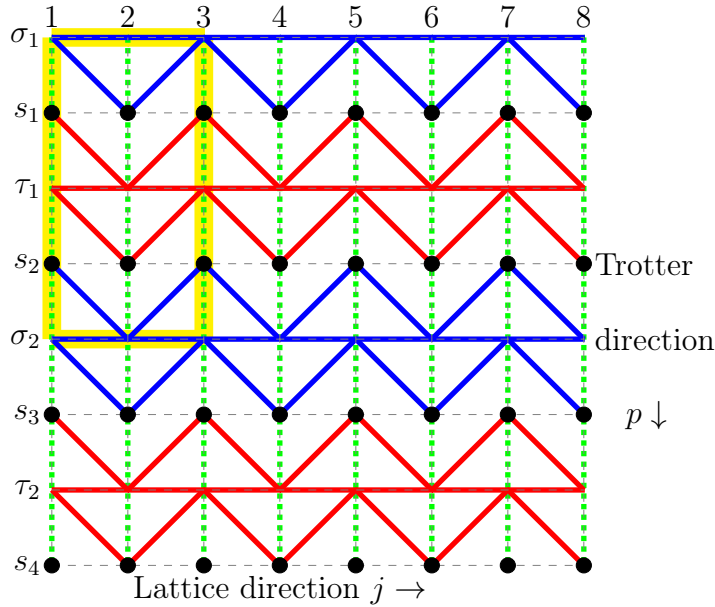
\begin{figure}[H]
  \centering
  \begin{tikzpicture}

    
		\node [left] at (1,8) {$\sigma_{1}$};
		\node [left] at (1,7) {$s_{1}$};
		\node [left] at (1,6) {$\tau_{1}$};
		\node [left] at (1,5) {$s_{2}$};
		\node [left] at (1,4) {$\sigma_{2}$};
		\node [left] at (1,3) {$s_{3}$};
		\node [left] at (1,2) {$\tau_{2}$};
		\node [left] at (1,1) {$s_{4}$};
		
		\node [below] at (4,1) {Lattice direction $j \rightarrow$};
		\node [right] at (8,5)  {Trotter};
		\node [right] at (8,4)  {direction};
		\node [right] at (8,3)  {$\quad p \downarrow$};
		\node [above] at (1,8) {1};
		\node [above] at (2,8) {2};
		\node [above] at (3,8) {3};
		\node [above] at (4,8) {4};
		\node [above] at (5,8) {5};
		\node [above] at (6,8) {6};
		\node [above] at (7,8) {7};
		\node [above] at (8,8) {8};
		
		\draw[line width = 7,yellow] (1,8) grid[step=1cm] (3,8);
		\draw[line width = 7,yellow] (1,4) grid[step=1cm] (3,4);
		\draw[line width = 7,yellow] (1,4) grid[step=1cm] (1,8);
		\draw[line width = 7,yellow] (3,4) grid[step=1cm] (3,8);
		
		\draw[style=dotted, line width = 2, green] (1,8) grid[step=1cm] (1,1);
		\draw[style=dotted, line width = 2, green] (2,8) grid[step=1cm] (2,1);
		\draw[style=dotted, line width = 2, green] (3,8) grid[step=1cm] (3,1);
		\draw[style=dotted, line width = 2, green] (4,8) grid[step=1cm] (4,1);
		\draw[style=dotted, line width = 2, green] (5,8) grid[step=1cm] (5,1);
		\draw[style=dotted, line width = 2, green] (6,8) grid[step=1cm] (6,1);
		\draw[style=dotted, line width = 2, green] (7,8) grid[step=1cm] (7,1);
		\draw[style=dotted, line width = 2, green] (8,8) grid[step=1cm] (8,1);


		\draw[line width = 2,blue] (1,8) grid[step=1cm] (8,8);
		\draw[line width = 2,blue] (1,4) grid[step=1cm] (8,4);
	
		\draw[line width = 2,red] (1,6) grid[step=1cm] (8,6);
		\draw[line width = 2,red] (1,2) grid[step=1cm] (8,2);

		\draw[line width = 2, blue] (1,8) -- (2,7);
		\draw[line width = 2, blue] (2,7) -- (3,8);
		\draw[line width = 2, blue] (3,8) -- (4,7);
		\draw[line width = 2, blue] (4,7) -- (5,8);
		\draw[line width = 2, blue] (5,8) -- (6,7);
		\draw[line width = 2, blue] (6,7) -- (7,8);
		\draw[line width = 2, blue] (7,8) -- (8,7);
		
		\draw[line width = 2, blue] (1,5) -- (2,4);
		\draw[line width = 2, blue] (2,4) -- (3,5);
		\draw[line width = 2, blue] (3,5) -- (4,4);
		\draw[line width = 2, blue] (4,4) -- (5,5);
		\draw[line width = 2, blue] (5,5) -- (6,4);
		\draw[line width = 2, blue] (6,4) -- (7,5);
		\draw[line width = 2, blue] (7,5) -- (8,4);

		\draw[line width = 2, blue] (1,4) -- (2,3);
		\draw[line width = 2, blue] (2,3) -- (3,4);
		\draw[line width = 2, blue] (3,4) -- (4,3);
		\draw[line width = 2, blue] (4,3) -- (5,4);
		\draw[line width = 2, blue] (5,4) -- (6,3);
		\draw[line width = 2, blue] (6,3) -- (7,4);
		\draw[line width = 2, blue] (7,4) -- (8,3);

		\draw[line width = 2, red] (1,7) -- (2,6);
		\draw[line width = 2, red] (2,6) -- (3,7);
		\draw[line width = 2, red] (3,7) -- (4,6);
		\draw[line width = 2, red] (4,6) -- (5,7);
		\draw[line width = 2, red] (5,7) -- (6,6);
		\draw[line width = 2, red] (6,6) -- (7,7);
		\draw[line width = 2, red] (7,7) -- (8,6);
		
		\draw[line width = 2, red] (1,6) -- (2,5);
		\draw[line width = 2, red] (2,5) -- (3,6);
		\draw[line width = 2, red] (3,6) -- (4,5);
		\draw[line width = 2, red] (4,5) -- (5,6);
		\draw[line width = 2, red] (5,6) -- (6,5);
		\draw[line width = 2, red] (6,5) -- (7,6);
		\draw[line width = 2, red] (7,6) -- (8,5);
		
		\draw[line width = 2, red] (1,3) -- (2,2);
		\draw[line width = 2, red] (2,2) -- (3,3);
		\draw[line width = 2, red] (3,3) -- (4,2);
		\draw[line width = 2, red] (4,2) -- (5,3);
		\draw[line width = 2, red] (5,3) -- (6,2);
		\draw[line width = 2, red] (6,2) -- (7,3);
		\draw[line width = 2, red] (7,3) -- (8,2);
		
		\draw[line width = 2, red] (1,2) -- (2,1);
		\draw[line width = 2, red] (2,1) -- (3,2);
		\draw[line width = 2, red] (3,2) -- (4,1);
		\draw[line width = 2, red] (4,1) -- (5,2);
		\draw[line width = 2, red] (5,2) -- (6,1);
		\draw[line width = 2, red] (6,1) -- (7,2);
		\draw[line width = 2, red] (7,2) -- (8,1);
		
    \draw[style=help lines,dashed] (1,1) grid[step=1cm] (8,8);
    \foreach \x in {1,...,8}{
      \foreach \y in {1,3,5,7}{
        \node[draw,circle,inner sep=2pt,fill] at (\x,\y) {};
      }
    }

  \end{tikzpicture}
  \caption{Lattice representation of a class of classical systems equivalent to the class of quantum systems \eqref{generalh} restricted to nearest and next nearest neighbours.}
  \label{figure:nnn}
\end{figure}

A schematic representation of this model on a two-dimensional lattice is given in Figure~\ref{figure:nnn}, with a yellow border representing a unit cell which can be repeated in either direction. The horizontal and diagonal blue and red lines represent interaction coefficients $J^{x}, J'^{x}$ and $J^{y}, J'^{y}$ respectively and the imaginary interaction coefficients by the dotted green lines. There is also a complex magnetic field term $ \left( \frac{\beta_{\mbox{\tiny qu}}}{2n} h \pm \frac{i \pi}{4} \right)$ applied to each site in every second row as represented by the black circles.

This mapping holds in the limit $n \rightarrow \infty$, which would result in coupling parameters $\frac{\beta_{\mbox{\tiny qu}}}{n} J^{x}, \frac{\beta_{\mbox{\tiny qu}}}{n} J^{y}, \frac{\beta_{\mbox{\tiny qu}}}{n} J'^{x}, \frac{\beta_{\mbox{\tiny qu}}}{n} J'^{y}, \frac{\beta_{\mbox{\tiny qu}} }{n} h \rightarrow 0$ unless we also take $\beta_{\mbox{\tiny qu}} \rightarrow \infty$. Therefore this gives us a connection between the ground state properties of the class of quantum systems and the finite temperature properties of the classical. 

Similarly to the nearest neighbour case, the partition function for this extended class of quantum systems can also be mapped to a class of classical vertex models (as we saw for the nearest neighbour case in Section~\ref{sec:nn}) or a class of classical models with up to 6-spin interactions around a plaquette with some extra constraints applied to the model (as we saw for the nearest neighbour case in Section~\ref{sec:nn}). We will not give the derivation of these as they are quite cumbersome and follow the same steps as outlined previously for the nearest neighbour cases, and instead include only the schematic representations of possible equivalent classical lattices. The interested reader can find the explicit computations in \cite{Jothesis}.

\begin{figure}[H]
  \centering
  \begin{tikzpicture}


  \draw [ultra thick, draw=black, fill=black!20!white]  (1,10) rectangle (3,9);
	\draw [ultra thick, draw=black, fill=black!20!white]  (4,10) rectangle (6,9);
	\draw [ultra thick, draw=black, fill=black!20!white]  (7,10) rectangle (9,9);
	\draw [ultra thick, draw=black, fill=black!20!white]  (7,10) rectangle (8,9);
	
	\draw [ultra thick, draw=black, fill=black!20!white]  (2,9) rectangle (4,8);
	\draw [ultra thick, draw=black, fill=black!20!white]  (5,9) rectangle (7,8);
	\draw [ultra thick, draw=black, fill=black!20!white]  (8,9) rectangle (9,8);
	
	\draw [ultra thick, draw=black, fill=black!20!white]  (1,8) rectangle (2,7);
	\draw [ultra thick, draw=black, fill=black!20!white]  (3,8) rectangle (5,7);
	\draw [ultra thick, draw=black, fill=black!20!white]  (6,8) rectangle (8,7);
	
	\draw [ultra thick, draw=black, fill=black!20!white]  (1,7) rectangle (3,6);
	\draw [ultra thick, draw=black, fill=black!20!white]  (4,7) rectangle (6,6);
	\draw [ultra thick, draw=black, fill=black!20!white]  (7,7) rectangle (9,6);
	\draw [ultra thick, draw=black, fill=black!20!white]  (7,7) rectangle (8,6);
	
	\draw [ultra thick, draw=black, fill=black!20!white]  (2,6) rectangle (4,5);
	\draw [ultra thick, draw=black, fill=black!20!white]  (5,6) rectangle (7,5);
	\draw [ultra thick, draw=black, fill=black!20!white]  (8,6) rectangle (9,5);
	
	\draw [ultra thick, draw=black, fill=black!20!white]  (1,5) rectangle (2,4);
	\draw [ultra thick, draw=black, fill=black!20!white]  (3,5) rectangle (5,4);
	\draw [ultra thick, draw=black, fill=black!20!white]  (6,5) rectangle (8,4);
	
	\draw [ultra thick, draw=black, fill=black!20!white]  (1,4) rectangle (3,3);
	\draw [ultra thick, draw=black, fill=black!20!white]  (4,4) rectangle (6,3);
	\draw [ultra thick, draw=black, fill=black!20!white]  (7,4) rectangle (9,3);
	\draw [ultra thick, draw=black, fill=black!20!white]  (7,4) rectangle (8,3);
	
	\draw [ultra thick, draw=black, fill=black!20!white]  (2,3) rectangle (4,2);
	\draw [ultra thick, draw=black, fill=black!20!white]  (5,3) rectangle (7,2);
	\draw [ultra thick, draw=black, fill=black!20!white]  (8,3) rectangle (9,2);
	
	\draw [ultra thick, draw=black, fill=black!20!white]  (1,2) rectangle (2,1);
	\draw [ultra thick, draw=black, fill=black!20!white]  (3,2) rectangle (5,1);
	\draw [ultra thick, draw=black, fill=black!20!white]  (6,2) rectangle (8,1);

		\node [left] at (1,10) {$s_{1}$};
		\node [left] at (1,9) {$s_{2}$};
		\node [left] at (1,8) {$s_{3}$};
		\node [left] at (1,7) {$s_{4}$};
		\node [left] at (1,6) {$s_{5}$};
		\node [left] at (1,5) {$s_{6}$};
		\node [left] at (1,4) {$s_{7}$};
		\node [left] at (1,3) {$s_{8}$};
		\node [left] at (1,2) {$s_{9}$};
		\node [left] at (1,1) {$s_{10}$};
		
		\node [below] at (5,1) {Lattice direction $j \rightarrow$};
		\node [right] at (9,6)  {Trotter};
		\node [right] at (9,5)  {direction};
		\node [right] at (9,4)  {$\quad p \downarrow$};
		\node [above] at (1,10) {1};
		\node [above] at (2,10) {2};
		\node [above] at (3,10) {3};
		\node [above] at (4,10) {4};
		\node [above] at (5,10) {5};
		\node [above] at (6,10) {6};
		\node [above] at (7,10) {7};
		\node [above] at (8,10) {8};
		\node [above] at (9,10) {9};

    \draw[style=help lines,dashed] (1,1) grid[step=1cm] (9,10);
    \foreach \x in {1,...,9}{
      \foreach \y in {1,...,10}{
        \node[draw,circle,inner sep=2pt,fill] at (\x,\y) {};
      }
    }

  \end{tikzpicture}
  \caption{Lattice representation of a class of classical systems equivalent to the class of quantum systems \eqref{generalh} restricted to nearest and next nearest neighbour interactions. The shaded areas indicate which particles interact together. }
  \label{fig:nnn9}
\end{figure}
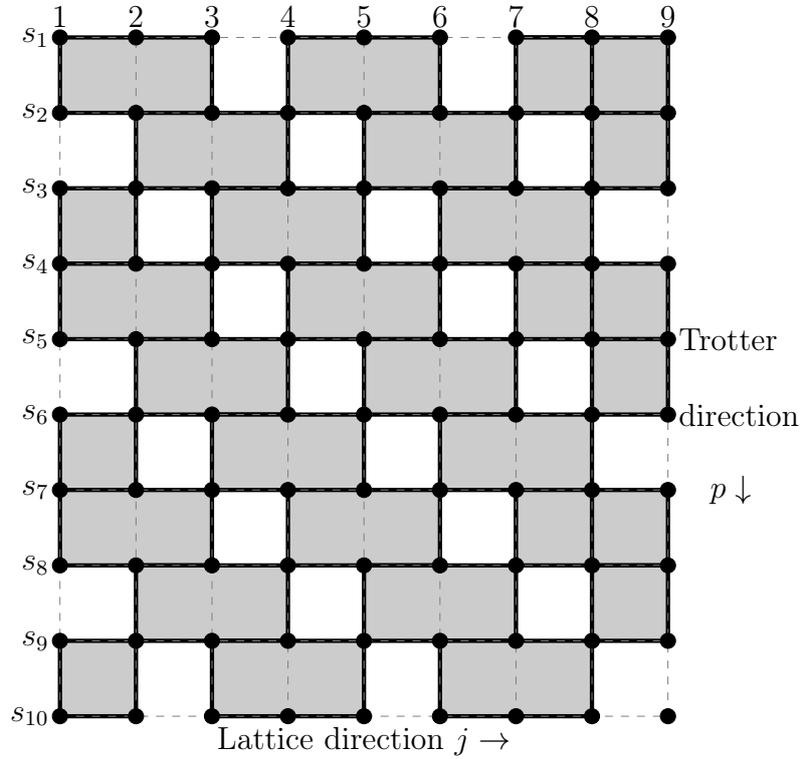

Firstly in Figure~\ref{fig:nnn9} we present a schematic representation of the latter of these two interpretations, a two-dimensional lattice of spins which interact with up to 6 other spins around the plaquettes shaded in grey.

\begin{figure}[H]
  \centering
  \begin{tikzpicture}


	\draw [ultra thick, draw=black, fill=black!20!white]  (2,2) rectangle (4,3);
	\draw [ultra thick, draw=black, fill=black!20!white]  (5,2) rectangle (7,3);

		\draw[line width = 2, blue] (2,2) -- (4,3);
		\draw[line width = 2, blue] (2,3) -- (4,2);
	
		\draw[line width = 2, blue] (5,2) -- (6,3);
		\draw[line width = 2, blue] (5,3) -- (6,2);
		\draw[line width = 2, blue] (6,2) -- (7,3);
		\draw[line width = 2, blue] (6,3) -- (7,2);
				
    \draw[style=help lines,dashed] (1,1) grid[step=1cm] (8,4);
    \foreach \x in {1,...,8}{
      \foreach \y in {1,...,4}{
        \node[draw,circle,inner sep=2pt,fill] at (\x,\y) {};
      }
    }

  \end{tikzpicture}
  \caption{Possible vertex representations}
  \label{fig:nnn99}
\end{figure}
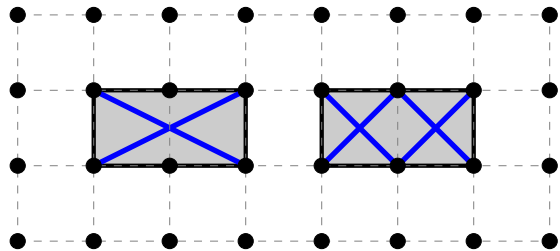

To imagine what the corresponding vertex models would look like, picture a line protruding from the lattice points bordering the shaded region and meeting in the middle of it. A schematic representation of two possible options for this is shown in Figure~\ref{fig:nnn99}. 

\subsection{Long range interactions}

For completeness we include the description of a classical system obtained by applying the Trotter-Suzuki mapping to the partition function for the general class of quantum systems \eqref{generalh} without any restrictions. 

We can now apply the Trotter expansion \eqref{tpf} to the quantum partition function with operators in the Hamiltonian \eqref{generalhpauli} ordered as
\begin{equation}
\begin{aligned}
Z = \lim_{n \rightarrow \infty} \mbox{Tr } [ & \prod^{M-1}_{j=1} (e^{\frac{\beta_{\mbox{\tiny qu}}}{n} \hat{\mathcal{H}}^{x}_{j,j+1}} e^{\frac{\beta_{\mbox{\tiny qu}}}{n} \hat{\mathcal{H}}^{x}_{j,j+2}} \ldots \\
& \ldots e^{\frac{\beta_{\mbox{\tiny qu}}}{n} \hat{\mathcal{H}}^{x}_{j,M}} e^{\frac{\beta_{\mbox{\tiny qu}}}{2n(M-1)} \hat{\mathcal{H}}^{z}} e^{\frac{\beta_{\mbox{\tiny qu}}}{n} \hat{\mathcal{H}}^{y}_{j,M}} \ldots e^{\frac{\beta_{\mbox{\tiny qu}}}{n} \hat{\mathcal{H}}^{y}_{j,j+2}} e^{\frac{\beta_{\mbox{\tiny qu}}}{n} \hat{\mathcal{H}}^{y}_{j,j+1}} ) ]^{n} \\
= \lim_{n \rightarrow \infty} \mbox{Tr } [ & \prod^{M}_{j=1} ( (\prod^{M-j}_{k=1} e^{\frac{\beta_{\mbox{\tiny qu}}}{n} \hat{\mathcal{H}}^{x}_{j,j+k}} ) e^{\frac{\beta_{\mbox{\tiny qu}}}{2n(M-1)} \hat{\mathcal{H}}^{z}} ( \prod^{M-j-1}_{k=0} e^{\frac{\beta_{\mbox{\tiny qu}}}{n} \hat{\mathcal{H}}^{y}_{j,M-k}} ) ) ]^{n},  
\label{z1gen}
\end{aligned}
\end{equation}
where $\hat{\mathcal{H}}^{\mu}_{j,k}  = J^{\mu}_{j,k} \sigma^{\mu}_{j} \sigma^{\mu}_{k} \prod^{k-1}_{l=1}\left(-\sigma^{z}_{l} \right)$ for $\mu \in x,y$ and $\hat{\mathcal{H}}^{z} = h \sum^{M}_{j=1} \sigma^{z}_{j}$.

For this model we need to insert $3Mn$ identity operators, $nM$ in each of the $\sigma^{x}$, $\sigma^{y}$ and $\sigma^{z}$ bases into \eqref{z1gen} in the following way;
\begin{equation}
\begin{aligned}
Z = &  \lim_{n \rightarrow \infty} \mbox{Tr } [ \prod^{M-1}_{j=1} ( \mathbb{I}_{\sigma_{j}} \left(\prod^{M-j}_{k=1} e^{\frac{\beta_{\mbox{\tiny qu}}}{n} \hat{\mathcal{H}}^{x}_{j,j+k}} \right) e^{\frac{\beta_{\mbox{\tiny qu}}}{(M-1)n} \hat{\mathcal{H}}^{z}}  \mathbb{I}_{s_{j}}  \\
& \qquad \times \left(\prod^{M-j-1}_{k=0} e^{\frac{\beta_{\mbox{\tiny qu}}}{n} \hat{\mathcal{H}}^{y}_{j,M-k}} \right) \mathbb{I}_{\tau_{j}} ) ]^{n} \\
= & \lim_{n \rightarrow \infty} \sum_{\sigma_{j,p}, \tau_{j,p}} \prod^{n-1}_{p=0} \prod^{M-1}_{j=1}  (\bra{\vec{\sigma}_{j+jp}} \left(\prod^{M-j}_{k=1} e^{\frac{\beta_{\mbox{\tiny qu}}}{n} \hat{\mathcal{H}}^{x}_{j,j+k}} \right) e^{\frac{\beta_{\mbox{\tiny qu}}}{n(M-1)} \hat{\mathcal{H}}^{z}}  \ket{\vec{s}_{j+jp}} \\
& \qquad \bra{\vec{s}_{j+jp}} \left( \prod^{M-j-1}_{k=0} e^{\frac{\beta_{\mbox{\tiny qu}}}{n} \hat{\mathcal{H}}^{y}_{j,M-k}}  \right) \ket{\vec{\tau}_{j+jp}} \braket{\vec{\tau}_{j+jp}}{\vec{\sigma}_{j+jp+1}}).
\label{z2gen}
\end{aligned}
\end{equation}

For this system it is then possible to rewrite the remaining matrix elements in \eqref{z2gen} in complex scalar exponential form by first writing
\begin{equation}
\begin{aligned}
& \bra{\vec{\sigma}_{j+jp}} \left(\prod^{M-j}_{k=1} e^{\frac{\beta_{\mbox{\tiny qu}}}{n} \hat{\mathcal{H}}^{x}_{j,j+k}} \right) e^{\frac{\beta_{\mbox{\tiny qu}}}{n(M-1)} \hat{\mathcal{H}}^{z}}  \ket{\vec{s}_{j+jp}}\\
&  \bra{\vec{s}_{j+jp}} \left(\prod^{M-j-1}_{k=0} e^{\frac{\beta_{\mbox{\tiny qu}}}{n} \hat{\mathcal{H}}^{y}_{j,M-k}} \right)   \ket{\vec{\tau}_{j+jp}} \braket{\vec{\tau}_{j+jp}}{\vec{\sigma}_{j+jp+1}}\\
& \quad =  e^{\frac{\beta_{\mbox{\tiny qu}}}{n} \sum^{M-j}_{k=1} \left(\mathcal{H}^{x}_{j,j+k}(p) + \mathcal{H}^{y}_{j,j+k}(p) + \frac{1}{n (M-1)} \mathcal{H}^{z}\right)} \braket{\vec{\sigma}_{j+jp}}{\vec{s}_{j+jp}}\\
& \qquad  \braket{\vec{s}_{j+jp}}{\vec{\tau}_{j+jp}} \braket{\vec{\tau}_{j+jp}}{\vec{\sigma}_{j+jp+1}}
\end{aligned}
\end{equation}
where $\mathcal{H}^{x}_{j,k}(p) =  \sum^{M}_{k=j+1} J^{x}_{j,k} \sigma_{j,p} \sigma_{k,p} \prod^{k-1}_{l=j+1} \left(-s_{l,p}\right)$, \\
$\mathcal{H}^{y}_{j,k}(p) = \sum^{M}_{k=j+1} J^{y}_{j,k} \tau_{j,p} \tau_{k,p} \prod^{k-1}_{l=j+1} \left(-s_{l,p}\right)$ and\\
$\mathcal{H}^{z}_{p} = h \sum^{M}_{j=1} \sigma^{z}_{j,p}$. Finally evaluating the remaining terms as
\begin{equation}
\begin{aligned}
\braket{\vec{\sigma}_{p}}{\vec{s}_{p}} & \braket{\vec{s}_{p}}{\vec{\tau}_{p}} \braket{\vec{\tau}_{p}}{\vec{\sigma}_{p+1}} \\
& \quad = \left(\frac{1}{2\sqrt{2}} \right)^{M} \prod^{M}_{j=1} e^{\frac{i \pi}{4} \left(\left(1-\sigma_{j,p} \right)\left(1-s_{j,p} \right)+\tau_{j,p} \left(1-s_{j,p}\right) -\sigma_{j,p+1} \tau_{j,p}\right) }.
\end{aligned}
\end{equation}

The partition function now has the same form as that of a class of two-dimensional classical Ising models on a $M \times 3Mn$ lattice with classical Hamiltonian $\mathcal{H}_{\mbox{\tiny cl}}$ given by
\begin{equation}
\begin{aligned}
& - \beta_{\mbox{\tiny cl}} \mathcal{H}_{\mbox{\tiny cl}} =  \sum^{n-1}_{p =1} \sum^{M}_{j=1} ( \frac{\beta_{\mbox{\tiny qu}}}{n} \sum^{M}_{k=j+1} (J^{x}_{j,k} \sigma_{j,j+jp} \sigma_{k,j+jp} \\
& + J^{y}_{j,k} \tau_{j,j+jp} \tau_{k,j+jp}  ) \prod^{k-1}_{l=j+1} \left(-s_{l,p}\right) +  \left(\frac{\beta_{\mbox{\tiny qu}}}{n(M-1)} h - \frac{i \pi}{4} \right)s_{j,j+jp} \\
& + \frac{i \pi}{4} ( 1-\sigma_{j,j+jp} + \tau_{j,j+jp} + \sigma_{j,j+jp} s_{j,j+jp} \\
& - \tau_{j,j+jp} s_{j,j+jp}  - \sigma_{j,j+jp+1}\tau_{j,j+jp} ) ) +  nM^{2} \ln \frac{1}{2 \sqrt{2}} .
\end{aligned}
\end{equation}

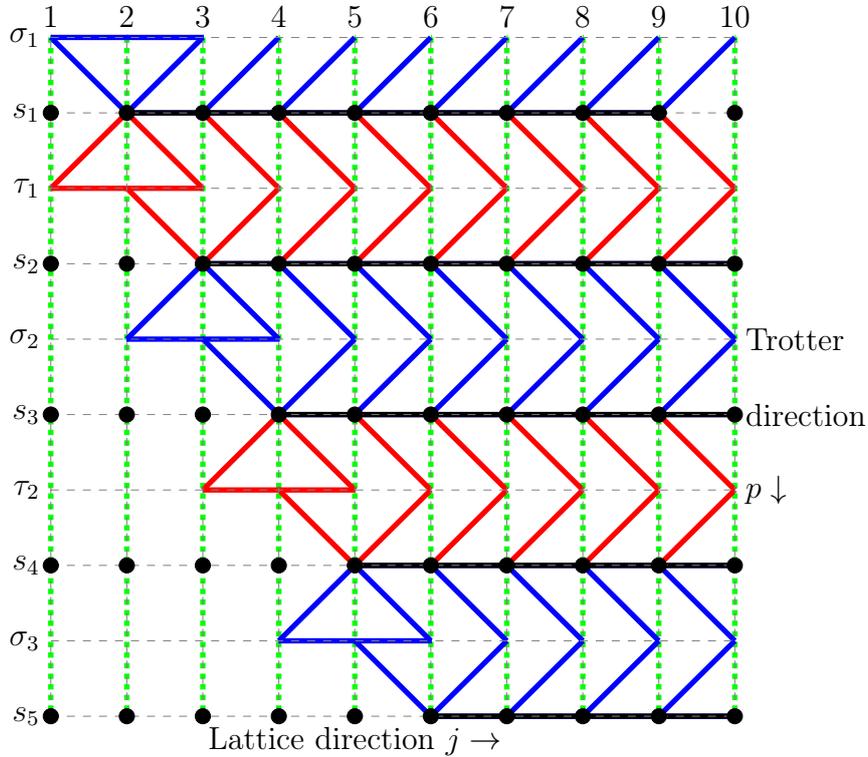
\begin{figure}[H]
  \centering
  \begin{tikzpicture}

    
		\node [left] at (1,10) {$\sigma_{1}$};
		\node [left] at (1,9) {$s_{1}$};
		\node [left] at (1,8) {$\tau_{1}$};
		\node [left] at (1,7) {$s_{2}$};
		\node [left] at (1,6) {$\sigma_{2}$};
		\node [left] at (1,5) {$s_{3}$};
		\node [left] at (1,4) {$\tau_{2}$};
		\node [left] at (1,3) {$s_{4}$};
		\node [left] at (1,2) {$\sigma_{3}$};
		\node [left] at (1,1) {$s_{5}$};
		
		\node [below] at (5,1) {Lattice direction $j \rightarrow$};
		\node [right] at (10,6)  {Trotter};
		\node [right] at (10,5)  {direction};
		\node [right] at (10,4)  {$p \downarrow$};
		\node [above] at (1,10) {1};
		\node [above] at (2,10) {2};
		\node [above] at (3,10) {3};
		\node [above] at (4,10) {4};
		\node [above] at (5,10) {5};
		\node [above] at (6,10) {6};
		\node [above] at (7,10) {7};
		\node [above] at (8,10) {8};
		\node [above] at (9,10) {9};
		\node [above] at (10,10) {10};

		\draw[style=dotted, line width = 2, green] (1,10) grid[step=1cm] (1,1);
		\draw[style=dotted, line width = 2, green] (2,10) grid[step=1cm] (2,1);
		\draw[style=dotted, line width = 2, green] (3,10) grid[step=1cm] (3,1);
		\draw[style=dotted, line width = 2, green] (4,10) grid[step=1cm] (4,1);
		\draw[style=dotted, line width = 2, green] (5,10) grid[step=1cm] (5,1);
		\draw[style=dotted, line width = 2, green] (6,10) grid[step=1cm] (6,1);
		\draw[style=dotted, line width = 2, green] (7,10) grid[step=1cm] (7,1);
		\draw[style=dotted, line width = 2, green] (8,10) grid[step=1cm] (8,1);
		\draw[style=dotted, line width = 2, green] (9,10) grid[step=1cm] (9,1);
		\draw[style=dotted, line width = 2, green] (10,10) grid[step=1cm] (10,1);

		\draw[line width = 2,blue] (1,10) grid[step=1cm] (3,10);
		\draw[line width = 2,blue] (3,7) grid[step=1cm] (10,7);
		\draw[line width = 2,blue] (4,5) grid[step=1cm] (10,5);   
		\draw[line width = 2,blue] (2,9) grid[step=1cm] (9,9); 
		\draw[line width = 2,blue] (2,6) grid[step=1cm] (4,6); 
		\draw[line width = 2,blue] (5,3) grid[step=1cm] (10,3);
		\draw[line width = 2,blue] (4,2) grid[step=1cm] (6,2);
		\draw[line width = 2,blue] (6,1) grid[step=1cm] (10,1);
			
		\draw[line width = 2,red] (1,8) grid[step=1cm] (3,8);
		\draw[line width = 2,red] (3,4) grid[step=1cm] (5,4);
		
		\draw[line width = 2,black] (2,9) grid[step=1cm] (9,9);
		\draw[line width = 2,black] (3,7) grid[step=1cm] (10,7);
		\draw[line width = 2,black] (4,5) grid[step=1cm] (10,5);
		\draw[line width = 2,black] (5,3) grid[step=1cm] (10,3);
		\draw[line width = 2,black] (6,1) grid[step=1cm] (10,1);

		\draw[line width = 2, blue] (1,10)--(2,9);
		\draw[line width = 2, blue] (2,9) -- (3,10);
		\draw[line width = 2, blue] (3,9) -- (4,10);
		\draw[line width = 2, blue] (4,9) -- (5,10);
		\draw[line width = 2, blue] (5,9) -- (6,10);
		\draw[line width = 2, blue] (6,9) -- (7,10);
		\draw[line width = 2, blue] (7,9) -- (8,10);
		\draw[line width = 2, blue] (8,9) -- (9,10);
		\draw[line width = 2, blue] (9,9) -- (10,10);
		
		\draw[line width = 2, blue] (2,6) -- (3,7);
		\draw[line width = 2, blue] (3,7) -- (4,6);
		\draw[line width = 2, blue] (4,7) -- (5,6);
		\draw[line width = 2, blue] (5,7) -- (6,6);
		\draw[line width = 2, blue] (6,7) -- (7,6);
		\draw[line width = 2, blue] (7,7) -- (8,6);
		\draw[line width = 2, blue] (8,7) -- (9,6);
		\draw[line width = 2, blue] (9,7) -- (10,6);
	
		\draw[line width = 2, blue] (3,6) -- (4,5);
		\draw[line width = 2, blue] (4,5) -- (5,6);
		\draw[line width = 2, blue] (5,5) -- (6,6);
		\draw[line width = 2, blue] (6,5) -- (7,6);
		\draw[line width = 2, blue] (7,5) -- (8,6);
		\draw[line width = 2, blue] (8,5) -- (9,6);
		\draw[line width = 2, blue] (9,5) -- (10,6);
		
		\draw[line width = 2, blue] (4,2) -- (5,3);
		\draw[line width = 2, blue] (5,3) -- (6,2);
		\draw[line width = 2, blue] (6,3) -- (7,2);
		\draw[line width = 2, blue] (7,3) -- (8,2);
		\draw[line width = 2, blue] (8,3) -- (9,2);
		\draw[line width = 2, blue] (9,3) -- (10,2);
	
		\draw[line width = 2, blue] (5,2) -- (6,1);
		\draw[line width = 2, blue] (6,1) -- (7,2);
		\draw[line width = 2, blue] (7,1) -- (8,2);
		\draw[line width = 2, blue] (8,1) -- (9,2);
		\draw[line width = 2, blue] (9,1) -- (10,2);

		\draw[line width = 2, red] (1,8) -- (2,9);
		\draw[line width = 2, red] (2,9) -- (3,8);
		\draw[line width = 2, red] (3,9) -- (4,8);
		\draw[line width = 2, red] (4,9) -- (5,8);
		\draw[line width = 2, red] (5,9) -- (6,8);
		\draw[line width = 2, red] (6,9) -- (7,8);
		\draw[line width = 2, red] (7,9) -- (8,8);
		\draw[line width = 2, red] (8,9) -- (9,8);
		\draw[line width = 2, red] (9,9) -- (10,8);
		
		\draw[line width = 2, red] (2,8) -- (3,7);
		\draw[line width = 2, red] (3,7) -- (4,8);
		\draw[line width = 2, red] (4,7) -- (5,8);
		\draw[line width = 2, red] (5,7) -- (6,8);
		\draw[line width = 2, red] (6,7) -- (7,8);
		\draw[line width = 2, red] (7,7) -- (8,8);
		\draw[line width = 2, red] (8,7) -- (9,8);
		\draw[line width = 2, red] (9,7) -- (10,8);
		
		\draw[line width = 2, red] (3,4) -- (4,5);
		\draw[line width = 2, red] (4,5) -- (5,4);
		\draw[line width = 2, red] (5,5) -- (6,4);
		\draw[line width = 2, red] (6,5) -- (7,4);
		\draw[line width = 2, red] (7,5) -- (8,4);
		\draw[line width = 2, red] (8,5) -- (9,4);
		\draw[line width = 2, red] (9,5) -- (10,4);
		
		\draw[line width = 2, red] (4,4) -- (5,3);
		\draw[line width = 2, red] (5,3) -- (6,4);
		\draw[line width = 2, red] (6,3) -- (7,4);
		\draw[line width = 2, red] (7,3) -- (8,4);
		\draw[line width = 2, red] (8,3) -- (9,4);
		\draw[line width = 2, red] (9,3) -- (10,4);

    \draw[style=help lines,dashed] (1,1) grid[step=1cm] (10,10);
    \foreach \x in {1,...,10}{
      \foreach \y in {1,3,5,7,9}{
        \node[draw,circle,inner sep=2pt,fill] at (\x,\y) {};
      }
    }

  \end{tikzpicture}
  \caption{Lattice representation of a classical system equivalent to the general class of quantum systems.}
  \label{figure:ghmethod2}
\end{figure}

A schematic representation of this class of classical systems on a two-dimensional lattice is given in Figure~\ref{figure:ghmethod2} where the blue and red lines represent interaction coefficients $J^{x}_{j,k}$ and $J^{y}_{j,k}$ respectively, the black lines are where they are both present and the imaginary interaction coefficients are given by the dotted green lines. The black circles also represent a complex field $ \left(\frac{\beta_{\mbox{\tiny qu}}}{n(M-1)} h - \frac{i \pi}{4} \right)$ acting on each individual particle in every second row.

This mapping holds in the limit $n \rightarrow \infty$, which would result in coupling parameters$\frac{\beta_{\mbox{\tiny qu}}}{n} J^{x}_{j,k}, \frac{\beta_{\mbox{\tiny qu}}}{n} J^{y}_{j,k}, \frac{\beta_{\mbox{\tiny qu}} }{n}h \rightarrow 0$ unless we also take $\beta_{\mbox{\tiny qu}} \rightarrow \infty$. Therefore this gives us a connection between the ground state properties of the quantum system and the finite temperature properties of the classical. 

\section{Systems equivalent to the dimer model.}
\label{sec:appapprelsvd}

We give here some explicit examples of relationships between parameters under which our general class of quantum spin chains \eqref{generalh} is equivalent to the two-dimensional classical dimer model using transfer matrix $\mathbf{V}^{2}_{d}$ \eqref{tmdim}.
\begin{itemize}
\item When $L=1$, from \eqref{tanis} we have
\begin{equation}
\begin{aligned}
- \frac{1}{\alpha \sin q}  & \neq \frac{b(1) \sin q}{\Gamma + a(1) \cos q}, \\
\end{aligned}
\end{equation}
therefore it is not possible to establish an equivalence in this case.

\item When $L=2$ from \eqref{tanis} we have
\begin{equation}
- \frac{1}{\alpha \sin q}   = \frac{ b(1)}{ - 2a(2) \sin q}, \quad \mbox{if } \Gamma=-a(2), \quad a(1)=b(2)=0,
\end{equation}
thus the systems are equivalent under the parameter relations
\begin{equation}
\alpha = \frac{2a(2)}{b(1)}, \quad \Gamma=-a(2), \quad a(1)=b(2)=0.
\end{equation}

\item When $L=3$ from \eqref{tanis} we have
\begin{equation}
\begin{aligned}
& - \frac{1}{\alpha \sin q} = - \frac{b(1) - b(3) + b(2) \cos q }{2 \sin q \left(a(2)+a(3)\cos q \right) }, \\
& \mbox{if } \Gamma =-a(2), \quad a(1)=-a(3), \quad b(3)=0,
\end{aligned}
\end{equation}
thus the systems are equivalent under the parameter relations
\begin{equation}
\begin{aligned}
\alpha = \frac{2a(3)}{b(2)}, \quad \frac{a(2)}{a(3)} = \frac{b(1)-b(3)}{b(2)}, \quad \Gamma =-a(2),\\
a(1)=-a(3), \quad b(3)=0.
\end{aligned}
\end{equation}
\end{itemize}

Therefore we find that in general when $L > 1$, we can use \eqref{tanis} to prove that we have an equivalence if
\begin{equation}
- \frac{1}{\alpha \sin q}  \\
= \frac{\sin q \sum^{m}_{k=1} b(k) \sum^{\left[\frac{k-1}{2}\right]}_{l=0} \binom{k}{2l+1} \sum^{l}_{i=0} \binom{l}{i} \left(-1\right)^{-i} \cos^{k-2i-1}q}{\Gamma+a(1) \cos q+\sum^{m}_{k=2} a(k) \sum^{\left[\frac{k}{2}\right]}_{l=0}(-1)^{l} \binom{k}{2l} \sin^{2l}q \cos^{k-2l}q}. 
\label{compfrac}
\end{equation}
We can write the sum in the denominator of \eqref{compfrac} as
\begin{equation}
\begin{aligned}
& \sum^{\left[\frac{m}{2}\right]}_{j=1} a(2j) + \cos q \sum^{\left[\frac{m}{2}\right]}_{j=1}  a(2j+1)  \\
& + \sin^{2}q( \sum^{\left[\frac{m}{2}\right]}_{j=1} a(2j)\sum^{j}_{i=1} \binom{j}{i}(-1)^{i} \sin^{2(i-1)}q  \\
& + \cos q \sum^{\left[\frac{m-1}{2}\right]}_{j=1}  a(2j+1)  \sum^{j}_{i=1} \binom{j}{i}(-1)^{i} \sin^{2(i-1)}q \\
& + \sum^{m}_{k=2} a(k) \sum^{\left[\frac{k}{2}\right]}_{l=1} (-1)^{l} \binom{k}{2l} \sin^{2(l-1)}q \cos^{k-2l}q). 
\label{den2}
\end{aligned}
\end{equation}

This gives us the following conditions
\begin{equation}
\begin{aligned}
\Gamma &= - \sum^{\left[\frac{m}{2}\right]}_{j=1} a(2j) \\
a(1) & = - \sum^{\left[\frac{m+1}{2}\right]}_{j=1} a(2j+1 ) = 0.
\end{aligned}
\end{equation}

We can then rewrite the remaining terms in the denominator \eqref{den2} as
\begin{equation}
\begin{aligned}
& \sin^{2} q ( \sum^{\left[\frac{m}{2}\right]}_{j=1} a(2j) \sum^{j}_{i=1} \binom{j}{i} \sum^{i-1}_{p=0} (-1)^{i+p} \cos^{2p}q \\
& + \sum^{\left[\frac{m-1}{2}\right]}_{j=1} a(2j+1) \sum^{j}_{i=1} \binom{j}{i} \sum^{i-1}_{p=0} (-1)^{i+p} \cos^{2p+1}q \\
& + \sum^{\left[\frac{m-1}{2}\right]}_{j=1} a(2j+1) \sum^{j}_{l=1} \binom{2j+1}{2l} \sum^{l-1}_{p=0}(-1)^{p+l} \cos^{2(j-p-1)+1}q )  \\
& + \sum^{\left[\frac{m}{2}\right]}_{j=1} a(2j) \sum^{j}_{l=1} \binom{2j}{2l} \sum^{l-1}_{p=0}(-1)^{p+l} \cos^{2(j-p-1)}q ). 
\label{den3}
\end{aligned}
\end{equation}

Finally we equate coefficients of matching powers of $\cos q$ in the numerator in \eqref{compfrac} and denominator \eqref{den3}. For example, this demands that $b(m) = 0$.

\end{appendices}

\end{document}